\documentclass[twocolumn,abs,table,dvipsnames]{aastex6}

\usepackage{graphicx}

\usepackage{aas_macros}
\usepackage{amsmath}
\usepackage{cases}
\usepackage{apjfonts}
\usepackage{tikz}
\usepackage{placeins}

\newcolumntype{L}[1]{>{\raggedright\arraybackslash}p{#1}}
\newcolumntype{C}[1]{>{\centering\arraybackslash}p{#1}}

\newcommand{\apjedit}[1]{#1}

\begin{document}

 \title{Mitigating complex dust foregrounds in future CMB polarization experiments}

\author{Brandon S. Hensley}
\author{Philip Bull}
\affiliation{Jet Propulsion Laboratory, California Institute of Technology, 4800 Oak Grove Drive, Pasadena, CA 91109, USA}
\email{brandon.s.hensley@jpl.nasa.gov}
\email{philbull@gmail.com}
\thanks{\copyright 2017 California Institute of Technology\\U.S. Government sponsorship acknowledged}

\date{\today}

\begin{abstract}
Polarized Galactic foregrounds are one of the primary sources of systematic error in measurements of the B-mode polarization of the Cosmic Microwave Background (CMB). Experiments are becoming increasingly sensitive to complexities in the foreground frequency spectra that are not captured by standard parametric models, potentially affecting our ability to efficiently separate out these components. Employing a suite of dust models encompassing a variety of physical effects, we simulate observations of a future seven-band CMB experiment to assess the impact of these complexities on parametric component separation. We identify configurations of frequency bands that minimize the `model errors' caused by fitting simple parametric models to more complex `true' foreground spectra, which bias the inferred CMB signal. We find that: (a) fits employing a simple two parameter modified blackbody (MBB) dust model tend to produce significant bias in the recovered polarized CMB signal in the presence of physically realistic dust foregrounds; (b) generalized MBB models with three additional parameters reduce this bias in most cases, but non-negligible biases can remain, and can be hard to detect; and (c) line of sight effects, which give rise to frequency decorrelation, and the presence of iron grains are the most problematic complexities in the dust emission for recovering the true CMB signal. More sophisticated simulations will be needed to demonstrate that future CMB experiments can successfully mitigate these more physically realistic dust foregrounds.
\end{abstract}


\section{Introduction}

B-mode polarization of the Cosmic Microwave Background (CMB) arises from gravitational lensing by large scale structure and, possibly, from gravitational waves generated during the inflationary epoch. Given the potential for constraining several important aspects of fundamental physics, such as the conditions of the very early Universe and the sum of the neutrino masses, the measurement of the B-mode signal is a subject of intense focus for current and planned ground, balloon, and space missions.

In addition to the cosmological signature, Galactic foregrounds such as dust and synchrotron emission are also able to produce B-mode polarization. Disentangling the Galactic and cosmological signatures is already a challenge for current experiments, which have placed upper limits on the tensor-to-scalar ratio $r$ of $\sim 0.1$. Achieving constraints on $r$ of order $10^{-3}$, as sought by proposed next generation experiments such as LiteBIRD \citep{LiteBIRD} and CMB-S4 \citep{CMBS4} will require subtraction of the Galactic signal with unprecedented accuracy.

Component separation in polarization can be significantly more complex than in total intensity. Polarized intensities add vectorially, with directions set by the interstellar magnetic field in the case of dust and synchrotron emission. Changes in the magnetic field direction along the line of sight coupled with spatial variations in the polarization spectra can result in different levels of cancellation/suppression of the polarized signal and a rotation of the polarization angle with frequency \citep[e.g.,][]{Tassis+Pavlidou_2015}. Thus, the observed polarized emission at one frequency is an imperfect predictor of the polarized emission at a different frequency, an effect termed `frequency decorrelation.' Frequency decorrelation effects, which arise whenever the foreground spectra have spatial variations, are expected to be present at some level, and some evidence for them has already been noted in the {\it Planck} data \citep{Planck_Int_L}. 

Frequency decorrelation poses a serious challenge for component separation methods; template-based methods can no longer rely on being able to factorize the frequency dependence and spatial variation of the foregrounds, while parametric spectral fitting methods require significantly more complex signal models \apjedit{\citep[e.g.,][]{Chluba+Hill+Abitbol_2017}} to account for the extra spectral structure that is induced.

Complexities in the emission physics of dust can also be amplified in polarization relative to total intensity. If the far-infrared (FIR) dust emission arises from two distinct dust components (e.g., silicate and carbonaceous grains) with different polarization properties, the total and polarized dust spectral energy distributions (SEDs) can differ significantly. For example, while both components will contribute to the total intensity signal, one may be significantly less polarized than the other, resulting in far weaker polarized emission. Likewise, magnetic dust grains may contribute negligibly in total intensity at frequencies higher than the microwave, but emit strong, polarized emission at lower frequencies \citep{Draine+Hensley_2013}. These scenarios are also challenging for component separation methods, which tend to assume relatively simple spectral models, and extrapolate foreground properties in the (much higher signal-to-noise) total intensity channel into polarization.

The risks of improper dust modeling are well-documented -- on intermediate angular scales, the residual foreground emission left after imperfect component separation can easily mimic the cosmological B-mode signal \citep{2014PhRvL.112x1101B, 2015PhRvL.114j1301B}, resulting in strongly biased cosmological parameter inferences. It is therefore critical to the success of future CMB polarization experiments that they can (a) model and separate Galactic dust emission over a wide range of possible emission physics scenarios; and (b) reliably identify situations in which the modeling and subtraction are inadequate, and hence may bias the recovery of the true CMB signal.

Some recent work has studied the ability of future experiments to reliably recover the polarized CMB signal in the face of complexities we consider here. \citet{ArmitageCaplan+etal_2012} found that neglecting the curvature in the dust SED due to departure from a pure Rayleigh-Jeans spectrum biased the recovered tensor-to-scalar ratio $r$ high by $\sim 1\sigma$. More recently, \citet{Remazeilles+etal_2016} evaluated the robustness of parametric component separation to multiple dust components for a suite of proposed CMB satellites. They found that fitting too simple a dust model was sufficient to bias $r$ by more than $5\sigma$ while maintaining an acceptable goodness of fit criterion, even for the most sensitive experiments. \apjedit{Similarly, \citet{Stompor+Errard+Poletti_2016} found that fitting a single dust component in the presence of multiple dust ``layers'' biased parametric component separation at the $\Delta r \sim 10^{-3}$ level.} \apjedit{\citet{Kogut+Fixsen_2016} considered dust SEDs with a distribution of dust temperatures as well as SEDs based on two-level systems, finding biases on $r$ of $\Delta r \sim 3\times10^{-3}$.} \citet{Poh+Dodelson_2017} evaluated the impact of multiple dust components along the line of sight aligned by different magnetic fields, finding that naive extrapolations from 350\,GHz to lower frequencies would result in significant bias for $r \lesssim 1.5\times 10^{-3}$.

In this paper, we consider the ability of future CMB polarization experiments to mitigate dust contamination in a much broader range of physically-motivated scenarios. We focus our investigation on parametric component separation methods, which employ physical models of the frequency-dependence of each emission component to perform separation pixel-by-pixel in the map domain \citep[e.g.,][]{Planck_2015_X}. This type of method produces maps of each foreground component in addition to the CMB, enabling a wide variety of Galactic science. It is also well-suited for cosmological applications at large angular scales, such as measuring the reionization peak, where techniques based on spatial correlations can fail due to lack of modes \citep{Remazeilles+etal_2017}. For higher $\ell$ applications ($\ell \gtrsim 100$), non-parametric techniques will still need to be tested for robustness against the model complexities discussed here.

We first assess which complications to the simplest models lead to the greatest biases in the recovered CMB. Guided by a physical understanding of dust emission, more sophisticated models and techniques can be developed to mitigate these biases. Second, by analyzing a large of set of frequency configurations that could be employed in future experiments, we evaluate what frequency coverage is most effective at both mitigating bias and identifying poor model fits via poor goodness of fit statistics.

This paper is organized as follows: in Section~\ref{sec:foreground}, we motivate and describe the suite of foreground models used in this work; in Section~\ref{sec:method}, we outline a ``single pixel'' (i.e., frequency-space) component separation method; we assess the ability of various mission designs to accurately recover the input CMB as a function of dust model complexity in Sections~\ref{sec:temp_pol} and \ref{sec:results}; and we discuss the implications of this analysis for experiment design and data analysis in Section~\ref{sec:discussion}.

\section{Foreground Models}
\label{sec:foreground}

In this section, we provide details of all of the foreground component models considered in this paper, with a particular focus on a set of seven dust models that illustrate a range of possible physical effects that would lead to more complex dust spectra than are typically considered in CMB analyses.

In most of this study, we will work in units of CMB brightness temperature. A blackbody of temperature $T_{\rm CMB}$ emits with a specific intensity of $I_\nu^{\rm CMB} = B_\nu\left(T_{\rm CMB}\right)$, 
where $B_\nu\left(T\right)$ is the Planck function and $T_{\rm CMB}$ is taken to be $2.7255$\,K. A source with specific intensity $I_\nu$ has a CMB brightness temperature $\Delta T$, satisfying:
\begin{equation}
\label{eq:tcmb}
I_\nu \simeq \left(\frac{\partial B_\nu}{\partial T}\right)_{T_{\rm CMB}}
\Delta T = \frac{2h\nu^3}{c^2} \frac{xe^x}{\left(e^x - 1\right)^2}
\frac{\Delta T}{T_{\rm CMB}}
~~~,
\end{equation}
where $x = h\nu/kT_{\rm CMB}$ and only the first order terms are retained. The foreground models presented in the following sections are expressed in terms of specific intensities and converted to CMB temperature units, $\mu{\rm K}_{\rm CMB}$, for the analysis.

It is often convenient, particularly for visualizing the SEDs of the emission components over a broad frequency range, to employ the Rayleigh-Jeans brightness temperature $T_{\rm RJ}$, which is related to the specific intensity at a given frequency by $T_{\rm RJ}(\nu) = {c^2\, I_\nu / (2 k\,\nu^2)}$.

For polarization, we work with the Stokes parameters $I_\nu$, $Q_\nu$, and $U_\nu$, expressed as specific intensities which can then be converted to the equivalent CMB or Rayleigh-Jeans brightness temperatures. Throughout this work, we neglect any circular polarization (i.e., Stokes $V_\nu = 0$). The polarized intensity $P_\nu$ therefore satisfies $P_\nu = |Q_\nu + iU_\nu|$. The polarization fraction $p_\nu$ is defined as $p_\nu \equiv P_\nu/I_\nu$.

\subsection{Component Amplitudes}
\label{sec:fg_amp}

\begin{figure*}
	\includegraphics[width=0.93\textwidth]{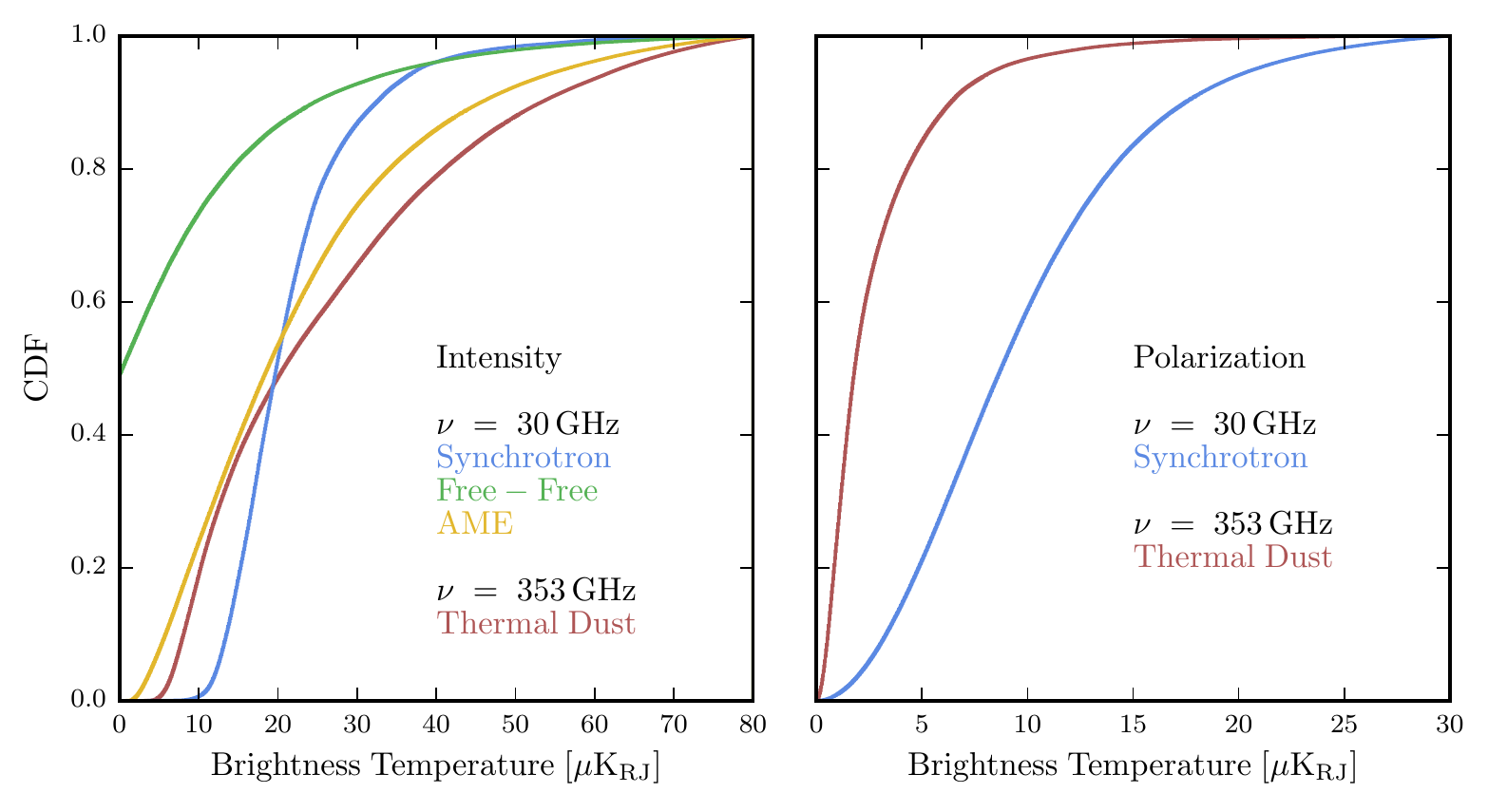}
	\caption{The cumulative distribution function (CDF) of total (left) and polarized (right) intensities for each foreground component in the $|b| > 30^\circ$ sky at $N_{\rm side} = 256$ as determined by the \texttt{Commander} analysis \citep{Planck_2015_X}. We employ these CDFs to select representative amplitudes for each component, which we summarize in Table~\ref{tbl:components}.}
	\label{fig:fiducial_amps}
\end{figure*}

\begin{table*}
 \centering
 \begin{tabular}{lccccl}
 \hline
 Component & Ref. freq. [GHz] & I [$\mu$K$_{\rm RJ}$] & Q [$\mu$K$_{\rm RJ}$] & U [$\mu$K$_{\rm RJ}$] & Spectral parameters \\
 \hline
 CMB & 30 & 50 & 0.6 & 0.6 & --- \\
 Synchrotron & 30 & 30 & 10 & 10 & $\beta_{\rm s} = -1.2$ \\
 Free-free & 30 & 30 & --- & --- & $\beta_{\rm ff} = -0.118$ \\
 AME & 30 & 30 & --- & --- & $\nu_{\rm pk} = 25$\,GHz \\
 Thermal Dust & 353 & 50 & 3.5 & 3.5 & Various \\
 \hline \\
 \end{tabular}
 \caption{Assumed amplitude and spectral parameters of the sky components in the simulations, based on the distributions from Planck shown in Figure~\ref{fig:fiducial_amps}.}
 \label{tbl:components}
\end{table*}

In the following sections, we describe models of the frequency dependence of the Galactic foregrounds considered in this work: thermal dust, synchrotron, anomalous microwave emission (AME), and free-free. As we implement a ``single pixel'' analysis in which we fit a single representative realization of the CMB and the foregrounds (see Section~\ref{sec:method}), we must first determine the relative amplitudes of the various components for a typical high latitude sightline.

To do so, we are guided by the results of the \texttt{Commander} component separation analysis \citep{Planck_2015_X}. In total intensity, the \texttt{Commander} analysis produced full sky maps at $1^\circ$ resolution pixellated at a \texttt{HealPix} \citep{Gorski+etal_2005} $N_{\rm side} = 256$.  In Figure~\ref{fig:fiducial_amps} we plot the cumulative distribution function (CDF) of the best fit (posterior maximum) amplitudes of thermal dust, synchrotron, free-free, and spinning dust (AME) for all pixels with $|b| > 30^\circ$. \apjedit{As the foreground brightness varies strongly with Galactic latitude, t}he highest values observed in this sky cut give an indication of the brightest foregrounds that need to be mitigated by an experiment wishing to achieve 50\% sky coverage.

At 353\,GHz, the distribution of posterior maximum dust intensities is $23_{-13}^{+30}\,\mu$K$_{\rm RJ}$ (68\% credible interval). For all of our dust models, we adopt a Stokes I amplitude of 50\,$\mu$K$_{\rm RJ}$ at 353\,GHz, which is at the higher end of this range.

At 30\,GHz, the distribution of best-fit synchrotron intensities is $19_{-5}^{+9}\,\mu$K$_{\rm RJ}$, and we adopt a value of 30\,$\mu$K$_{\rm RJ}$. Likewise, the AME amplitude distributions at 30\,GHz is $19_{-11}^{+21}\,\mu$K$_{\rm RJ}$ and we adopt a value of 30\,$\mu$K$_{\rm RJ}$. Free-free emission is not detected over a large fraction of the high latitude sky, as evidenced in Figure~\ref{fig:fiducial_amps}. 95\% of pixels have a 30\,GHz free-free amplitude less than 41\,$\mu$K$_{\rm RJ}$, and so we adopt an amplitude of 30\,$\mu$K$_{\rm RJ}$. Given the difficulty of separating synchrotron, AME, and free-free with the presently available low-frequency data, we purposefully model these components as equally bright at 30\,GHz in total intensity rather than adhere strictly to the \texttt{Commander} CDFs.

The \texttt{Commander} analysis also extended to polarization, producing full sky $N_{\rm side} = 256$ maps of polarized dust and synchrotron emission at $10'$ and $40'$ resolution respectively. We convert the \texttt{Commander} $Q_\nu$ and $U_\nu$ maps to $P_\nu$ and plot the resulting histogram over all high latitude ($|b| > 30^\circ$) pixels in Figure~\ref{fig:fiducial_amps}. The distribution of dust and synchrotron polarized intensities are $2.7_{-1.5}^{+3.1}$ and $8.9_{-4.6}^{+6.8}\,\mu$K$_{\rm RJ}$ at 353\,GHz and 30\,GHz respectively. Assuming a polarization angle of 22.5$^\circ$ for both dust and synchrotron, we adopt $Q_{\rm 353\,GHz} = U_{\rm 353\,GHz} = 3.5\,\mu$K$_{\rm RJ}$ and $Q_{\rm 30\,GHz} = U_{\rm 30\,GHz} = 10\,\mu$K$_{\rm RJ}$ for the dust and synchrotron, respectively. 

The \texttt{Commander} analysis assumed that the AME and free-free are unpolarized, following both theoretical and observational arguments which we outline in Sections~\ref{ssec:ame} and \ref{ssec:ff}. Likewise, we assume these components are unpolarized.

Finally, for the CMB we adopt a temperature of 50\,$\mu$K$_{\rm CMB}$ and Stokes $Q_\nu = U_\nu = 0.6\,\mu$K$_{\rm CMB}$, which are typical values for $\sim 1^\circ$ scales. The adopted amplitudes for all emission components are summarized in Table~\ref{tbl:components}.

While these choices result in a representative, benchmark realization of the foreground components, we note that the ratio of component amplitudes is observed to vary strongly across the sky \citep{Planck_2015_X}. Detailed optimization studies should explore a range of foreground realizations, which is beyond the scope of the present analysis.

\subsection{Thermal Dust}
\label{sec:thdust}

\begin{figure*}
	\includegraphics[width=\textwidth]{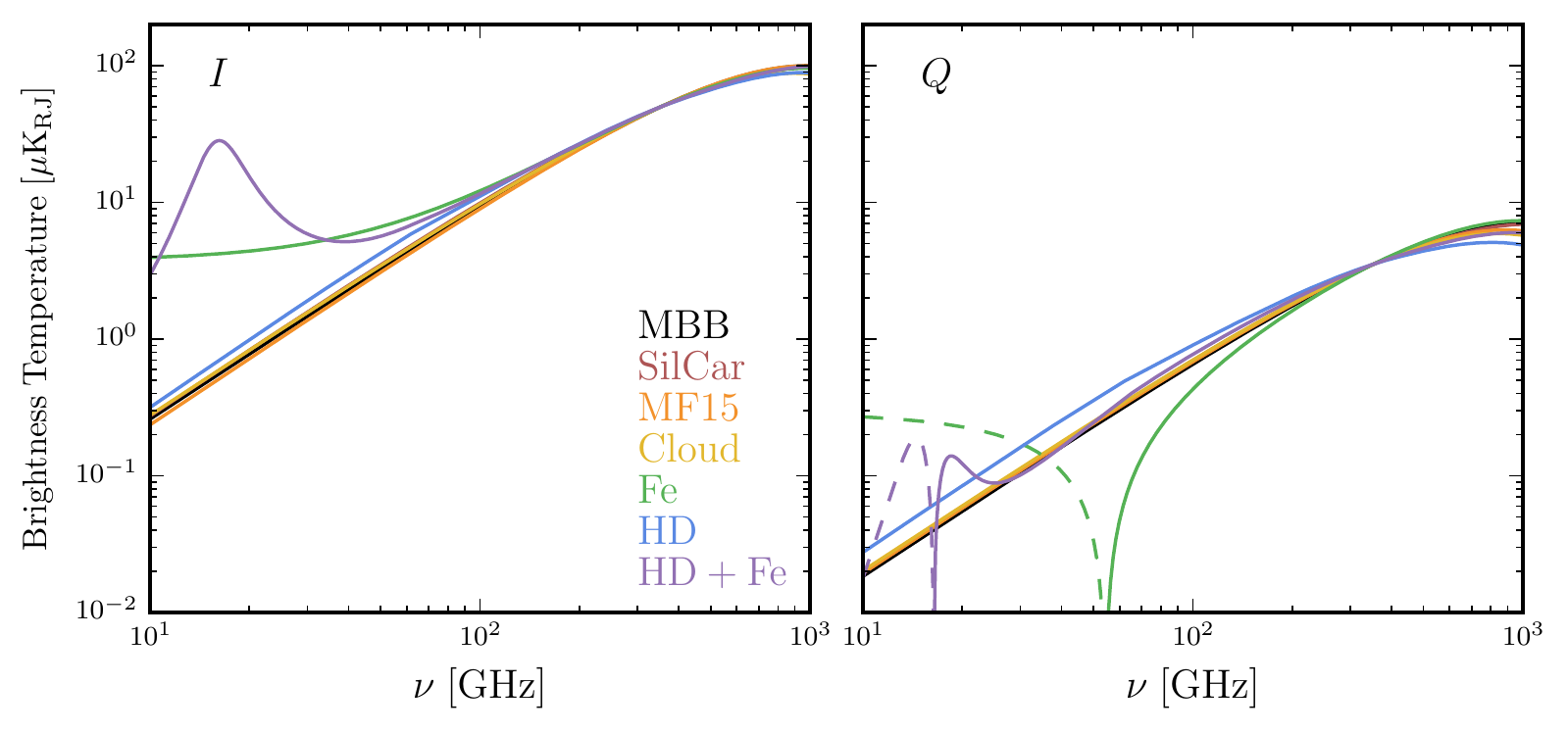}
	\caption{Input SEDs for the various dust models in Stokes I and Q. For most models, Stokes U is nearly identical to Q. The dashed lines indicate where Q is negative.}
	\label{fig:dust_models}
\end{figure*}

In this section, we define a suite of seven dust models that exhibit a range of complex but physically-motivated behaviors that could prove challenging for parametric component separation techniques. The dust models, and the parameters that characterize them, are summarized in Table~\ref{table:dust_models}, and representative SEDs are plotted in Figure~\ref{fig:dust_models}.

\subsubsection{Generalized Modified Blackbody}
\label{sec:gmbb}

Dust grains absorb optical and UV photons and reradiate the absorbed energy in the infrared. The far-infrared (FIR) opacity of dust grains is often approximated as a power law in frequency,
\begin{equation}
\kappa_\nu^d = \kappa_0 \left(\frac{\nu}{\nu_0}\right)^{\beta_d}
\end{equation}
where $\kappa_0$ is the opacity at reference frequency $\nu_0$. Assuming this opacity law, the total intensity emitted at frequency $\nu$ by a dust grain of temperature $T_d$ is
\begin{equation} \label{eq:sT_mbb_I}
I_\nu^d = A_d^I \left(\frac{\nu}{\nu_0^d}\right)^{\beta_d} B_\nu\left(T_d\right)
~~~,
\end{equation}
where $A_d^I$ is a dimensionless amplitude parameter and $B_\nu\left(T\right)$ is the Planck function. For convenience, we adopt $\nu_0^d = 353$\,GHz.

The polarization of the dust emission depends on the angle $\Psi$ between the interstellar magnetic field (along which the grains are aligned) and the line of sight, as well as the material composition and shape of the emitting grains. Variations in the magnetic field direction and grain properties along the line of sight can further influence the polarization signal.

For a single dust grain of temperature $T_d$ and ignoring any frequency-dependence of the polarization fraction arising from the frequency-dependence of its dielectric function, we can approximate the polarized intensity as
\begin{equation}
\label{eq:sT_mbb_P}
P_\nu^d = A_d^P \left(\frac{\nu}{\nu_0^d}\right)^{\beta_d}
B_\nu\left(T_d\right)
~~~,
\end{equation}
where $A_d^P$ is the specific polarized intensity at a reference frequency $\nu_0^d$. Note that the frequency-independent factor $\sin^2 \Psi$ is subsumed in the amplitude $A_d^P$.

As the Stokes parameters $Q_\nu$ and $U_\nu$ are more fundamental
observables than $P_\nu$, we construct the amplitudes
\begin{align}
A_d^Q &= A_d^P  \cos 2\gamma_d \\
A_d^U &= A_d^P  \sin 2\gamma_d
\end{align}
where $0 \leq \gamma_d < \pi$.

Let us now consider emission from an ensemble of grains. If there are $N$ distinct grain types, each with their own $\beta_d$, $T_d$, and polarization angle $\gamma_d$, then the total and polarized intensities from the ensemble are given by:
\begin{align}
I_\nu^d &= \sum_{j=1}^N A_{d,j}^I \left(\frac{\nu}{\nu_0^d}\right)^{\beta_{d,j}} B_\nu\left(T_{d,j}\right) \label{eq:G_mbb_I}\\
Q_\nu^d &= \sum_{j=1}^N A_{d,j}^Q \left(\frac{\nu}{\nu_0^d}\right)^{\beta_{d,j}}
B_\nu\left(T_{d,j}\right) \label{eq:G_mbb_Q} \\
U_\nu^d &= \sum_{j=1}^N A_{d,j}^U \left(\frac{\nu}{\nu_0^d}\right)^{\beta_{d,j}}
B_\nu\left(T_{d,j}\right)\label{eq:G_mbb_U} \\
\gamma_\nu^d &= \frac{1}{2}\,{\rm arg}\left(Q_\nu^d + iU_\nu^d\right)\label{eq:G_mbb_gamma}
~~~,
\end{align}
where arg denotes the principal value of the argument function with range $[0,2\pi)$. We denote this model as a ``generalized modified blackbody'' and employ it as a starting point for a number of simple analytic models which we discuss in the following sections.

\subsubsection{Single Component Modified Blackbody (MBB)}
\label{sec:mbb}
If all interstellar dust has the same $\beta_d$ (e.g., if all grains are of the same composition) and same temperature $T_d$ (e.g., if all grains are exposed to identical radiation fields), then the dust emission may be modeled as a single-temperature modified blackbody described by Equations~\ref{eq:G_mbb_I}--\ref{eq:G_mbb_U} with $N = 1$. The polarization angle is frequency independent.


When simulating dust emission with this model, we adopt $\beta_d = 1.6$ and $T_d = 20$\,K, consistent with typical values used to fit {\it Planck} data \citep{Planck_2013_XI,Planck_2015_X}. To achieve the adopted 353\,GHz total and polarized dust intensities, we take $A_d^I = 3.9\times10^{-6}$ and $A_d^Q = A_d^U = 2.8\times10^{-7}$.

Note that this model is the most commonly used in parametric dust component fits to CMB data, primarily due to its simplicity -- there are only two spectral parameters, $\beta_d$ and $T_d$, which are the same for the total intensity and polarized emission. We refer to this model as {\it MBB}.

\subsubsection{Two Component Modified Blackbody (2MBB)}
\label{sec:2mbb}
In this model we assume that dust comes in two distinct compositions (e.g., carbonaceous and silicate) with different $\beta_d$ and $T_d$. We further assume that these grains are aligned by the same magnetic field and therefore have the same polarization angle. Thus the total and polarized dust emission are given by Equations~\ref{eq:G_mbb_I}--\ref{eq:G_mbb_U} with $N = 2$, along with the constraint
\begin{equation}
{\rm arg}\left(A_{d,1}^Q + iA_{d,1}^U\right) = {\rm arg}\left(A_{d,2}^Q + iA_{d,2}^U\right)
\end{equation}
which enforces consistency between the polarization angles of the two components. The resulting polarization angle is therefore frequency-independent. 

When performing simulations with this model, we choose two sets of parameters. For the first set, we consider a physically motivated model intended to represent silicate and carbonaceous grains. We choose grain temperatures of $T_{d,1} = 15$\,K and $T_{d,2} = 24$\,K for silicate and carbonaceous grains, respectively, in agreement with the steady-state temperatures of $0.1\,\mu$m grains in the \citet{Hensley+Draine_2017c} model. Likewise, we employ spectral indices of $\beta_{d,1} = 1.6$ and $\beta_{d,2} = 1.8$. On the basis of both observational and theoretical results, we assume the carbonaceous grains are unaligned, and so have no net polarization \citep{Chiar+etal_2006,Hoang+Lazarian_2016}. Guided by the relative contributions of the silicate and carbonaceous grains to the total infrared emission in the \citet{Hensley+Draine_2017c} model, we adopt $f_I \equiv A_{d,2}^I/A_{d,1}^I = 0.25$, $f_Q \equiv A_{d,2}^Q/A_{d,1}^Q = 0$, and $f_U \equiv A_{d,2}^U/A_{d,1}^U = 0$, $A_{d,1}^I = 3.5\times10^{-6}$, and $A_{d,1}^Q = A_{d,1}^U = 3.3\times10^{-7}$ to achieve the desired total and polarized intensities. We refer to this model as {\it SilCar}.

For the second set, we employ the best-fit parameters of an empirical two component model developed to fit the observations of Galactic FIR dust emission. Both \citet{Finkbeiner+Davis+Schlegel_1999}, using FIRAS, IRAS, and DIRBE data, and more recently \citet{Meisner+Finkbeiner_2015}, using {\it Planck} data, found a statistically significant preference for two component models over one component models when fitting thermal dust emission. The best-fit parameters of the \citet{Meisner+Finkbeiner_2015} model are $T_{d,1} = 9.75$\,K, $T_{d,2} = 15.70$\,K, $\beta_{d,1} = 1.63$, $\beta_{d,2} = 2.82$, and $A_{d,1}^I/A_{d,2}^I = 5.35$. To achieve a 353\,GHz dust brightness temperature of 50\,$\mu$K$_{\rm RJ}$, we set $A_{d,1}^I = 9.5\times10^{-6}$ and $A_{d,2}^I = 1.8\times10^{-6}$. For polarization we adopt $A_{d,1}^Q = A_{d,1}^U = 6.7\times10^{-7}$ and $A_{d,2}^Q = A_{d,2}^U = 1.3\times10^{-7}$ to yield a total polarized intensity of $5\,\mu$K$_{\rm RJ}$ at 353\,GHz, and the same ratio between components as in total intensity. We refer to this model as {\it MF15}.

Note that we will sometimes use the definitions $\Delta \beta_d \equiv \beta_{d,2} - \beta_{d,1}$ and $f_X \equiv A_{d,2}^X / A_{d,1}^X$ (where $X = I, Q, U$) in our discussion of 2MBB models.

\subsubsection{``Cloud'' Model}
\label{sssec:mbb_cloud}

In the models discussed so far, the polarization angle has been constant with frequency. There are many scenarios in which this will not be the case, such as when both the magnetic field direction and the dust spectrum are varying along the line of sight.

\begin{table*}[t]
\begin{deluxetable*}{lccc}
      \tablecaption{Summary of Dust Models \label{table:dust_models}}
    \tablehead{\colhead{Model} & \colhead{Equations} & \colhead{Components} & \colhead{Constraints}}
    \startdata
Modified Blackbody & \ref{eq:G_mbb_I}--\ref{eq:G_mbb_U} & 1 &
--\\
Silicate + Carbonaceous & \ref{eq:G_mbb_I}--\ref{eq:G_mbb_U} & 2 & ${\rm arg}\left(A_{d,1}^Q + iA_{d,1}^U\right) = {\rm arg}\left(A_{d,2}^Q + iA_{d,2}^U\right)$ \\ 
MF15 & \ref{eq:G_mbb_I}--\ref{eq:G_mbb_U} & 2 & ${\rm arg}\left(A_{d,1}^Q + iA_{d,1}^U\right) = {\rm arg}\left(A_{d,2}^Q + iA_{d,2}^U\right)$ \\
Cloud & \ref{eq:G_mbb_I}--\ref{eq:G_mbb_U} & 2 & $\beta_{d,1} = \beta_{d,2}$ \\
Silicate + Fe & \ref{eq:G_mbb_I}--\ref{eq:G_mbb_U} & 2 & $\beta_{d,2} = 0$,\ $T_{d,1} = T_{d,2}$, \\
& & & ${\rm arg}\left(A_{d,1}^Q + iA_{d,1}^U\right) = -{\rm arg}\left(A_{d,2}^Q + iA_{d,2}^U\right)$\\
HD & \ref{eq:HD_I}--\ref{eq:HD_U} & 2 & --\\
HD with Fe & \ref{eq:HDFe_I}--\ref{eq:HDFe_U} & 2 & --\\
    \enddata
\end{deluxetable*}
\end{table*}

To model this effect, we consider the simple case in which there are two clouds along the line of sight. The dust in each cloud is identical in composition (i.e., has the same $\beta_d$), but is heated to different temperatures. Further, the clouds differ in magnetic field direction, yielding different polarization angles. In this case, the total and polarized intensities are given by Equations~\ref{eq:G_mbb_I}--\ref{eq:G_mbb_U} with $N = 2$ and $\beta_{d,1} = \beta_{d,2}$. Since the dust SEDs of the clouds are different, the polarization angle is frequency-dependent.

When simulating data with this model, we choose dust temperatures of $T_{d,1} = 15$\,K and $T_{d,2} = 20$\,K, well within the observed range of dust temperature variations \citep[e.g.][]{Planck_2013_XI,Meisner+Finkbeiner_2015,Planck_2015_X}. We assume that the two components are comparably bright in intensity, with $A_{d,1}^I = A_{d,2}^I = 2.4\times10^{-6}$. The adopted polarization amplitudes are $A_{d,1}^Q = 1.2\times10^{-7}$, $A_{d,1}^U = 2.1\times10^{-7}$, $A_{d,2}^Q = 2.4\times10^{-7}$, and $A_{d,1}^Q = 1.1\times10^{-7}$, corresponding to polarization angles of 30$^\circ$ and 12$^\circ$ for components 1 and 2, respectively. The magnetic fields in the clouds are therefore somewhat, but not extremely, misaligned, providing an indication for the typical magnitude of this effect. We refer to this model as {\it Cloud}.

\subsubsection{Modified Blackbody with Iron Grains (Fe)}
\label{ssec:mbb_fe}

Iron is a major constituent of interstellar dust by mass, some of which may be in the form of ferromagnetic nanoparticles \citep{Draine+Hensley_2013}. Indeed, embedded metallic iron nanoparticles have been found in putative interstellar grains collected in the Solar System by Stardust \citep{Westphal+etal_2014} and Cassini \citep{Altobelli+etal_2016}. The unique polarization signature of magnetic dipole emission potentially renders these grains an important contaminant for CMB studies -- the SED is relatively flat (in CMB temperature units) at low frequencies (see Figure~\ref{fig:dust_models}). We will consider the case of ferromagnetic inclusions embedded in larger non-magnetic grains. 

We can model the composite grains in the context of the generalized modified blackbody model. The non-magnetic matrix of the grain can be described by a simple single-temperature modified blackbody, while the iron inclusions can be modeled as a second modified blackbody of the same temperature but with $\beta = 0$ and polarization angle rotated by 90$^\circ$ with respect to the non-magnetic matrix \citep{Draine+Hensley_2013}. Thus, the total and polarized intensities for such composite grains are given by Equations~\ref{eq:G_mbb_I}--\ref{eq:G_mbb_U} with $N = 2$, $\beta_{d,2} = 0$, and
\begin{equation}
{\rm arg}\left(A_{d,1}^Q + iA_{d,1}^U\right) = -{\rm arg}\left(A_{d,2}^Q + iA_{d,2}^U\right)
~~~.
\end{equation}
We note that when magnetic dipole emission from the iron grains becomes larger than the electric dipole emission from the matrix, the polarization angle flips from $\gamma_{d,1}$ to $\gamma_{d,2} = \gamma_{d,1} + \pi/2$; the two types of emission are polarized perpendicular to one another.

When simulating data with this model, we employ $T_d = 20$\,K and $\beta_d = 1.6$, as the presence of iron inclusions do not substantially affect the dust temperature. At the reference frequency of 353\,GHz, we assume that the emission from the iron is 5\% of that of the silicate matrix in both intensity and polarization, and so $A_{d,2}^I/A_{d,1}^I = 0.05$ and $A_{d,2}^Q/A_{d,1}^Q = A_{d,2}^U/A_{d,1}^U = -0.05$. This provides a good approximation to the more physical model of \citet{Hensley+Draine_2017c} (see Figure~\ref{fig:dust_models}). To achieve the adopted 353\,GHz total and polarized dust intensities in Table~\ref{tbl:components}, we take $A_{d,1}^I = 3.7\times10^{-6}$ and $A_{d,1}^Q = A_{d,1}^U = 2.9\times10^{-7}$. We refer to this as the {\it Fe} model.

\subsubsection{A Physical Dust Model (HD and HD+Fe)}
\label{sec:hd16}

The analytic modified blackbody models presented in the previous sections simplify considerably the underlying dust physics. In this section we consider dust models that allow for much greater complexity and which may better reflect the challenges posed by the true dust foreground.

The modified blackbody model of dust emission assumes that the dust opacity $\kappa$ averaged over all dust sizes and compositions has a frequency-dependence that is well-described by a power law of variable index $\beta_d$. In a more physical treatment of dust, we compute the dust opacity as a function of grain size for various grain materials based on their complex dielectric function. We designate the opacity of a grain of composition $j$ and radius $a$ at frequency $\nu$ as $\kappa_{\nu, j, a}$.

In the modified blackbody model, the dust emission is characterized by a single temperature $T_d$. In reality, grains of different sizes and compositions will attain different temperatures even when exposed to the same radiation field. Further, very small grains are poorly described by a single temperature. A single optical or UV photon can excite these grains to temperatures exceeding 1000\,K, whereas most of the time the grain remains in its very cold ground vibrational state. Hence, it is important to consider the full temperature probability distribution for such grains. 

Accounting for these complexities, the total specific intensity from a population of dust grains is
\begin{equation}
\label{eq:phys_int}
I_\nu = \sum_j \int {\rm d}a \frac{{\rm d}m_i}{{\rm d}a} \int {\rm d}T
\left(\frac{{\rm d}P}{{\rm d}T}\right)_{\chi,j,a} \kappa_{\nu, j, a}
B_\nu\left(T\right)
~~~,
\end{equation}
where we have summed over all of the grain compositions $j$, (d$m_j$/d$a$)d$a$ is the mass in dust of composition $j$ with radius between $a$ and $a+$d$a$, $\chi$ is a parameter governing the strength of the radiation field heating the dust, and (d$P$/d$T$)d$T$ is the probability of a grain of composition $j$ and radius $a$ in a radiation field $\chi$ having temperature between $T$ and $T+$d$T$. The radiation field is assumed to be a scalar multiple $\chi$ of the spectrum of the local interstellar radiation field derived by \citet{Mathis+Mezger+Panagia_1983}, which we set to $\chi = 1$ here.

We employ Equation~\ref{eq:phys_int} directly in the context of a physical dust model. By analyzing dust emission and extinction, both total and polarized, from ultraviolet to microwave wavelengths, \citet{Hensley+Draine_2017c} developed a model based on graphitic and silicate grains capable of reproducing the observations. By adopting the \citet{Hensley+Draine_2017c} opacities, size distributions, and temperature distributions, we can reduce Equation~\ref{eq:phys_int} to just a few key parameters:
\begin{equation}
\label{eq:hd_int}
I_\nu = \sum_j A_{d,j}^I I_\nu\left(j, \chi\right)
~~~,
\end{equation}
where the $A_{d,j}^I$ set the relative contributions of the various grain components as well as the overall amplitude, and the $I_\nu\left(j, \chi\right)$ are precomputed quantities based on the physical modeling.

Just as the physical models provide precomputed $I_\nu\left(\chi\right)$ based on realistic grain materials, so too can they provide $P_\nu\left(\chi\right)$ as long as the grain shapes are specified. In addition to the intrinsic grain properties, the observed $Q_\nu$ and $U_\nu$ depend on the relative orientation of the interstellar magnetic field and both the line of sight and the reference polarization axes. We subsume these angles into the amplitude parameters $A_d^Q$ and $A_d^U$.

In this work, we explore two such physical dust models. In the first, we assume grains are either carbonaceous or silicate and with no embedded iron inclusions. The total and polarized intensities in this model are
\begin{align}
I_\nu^{\rm HD} &= A_{d,1}^I I_\nu^{\rm sil} + A_{d,2}^I I_\nu^{\rm car} \label{eq:HD_I} \\
Q_\nu^{\rm HD} &= A_{d,1}^Q Q_\nu^{\rm sil} + A_{d,2}^Q Q_\nu^{\rm car} \label{eq:HD_Q} \\
U_\nu^{\rm HD} &= A_{d,1}^U U_\nu^{\rm sil} + A_{d,2}^U U_\nu^{\rm car} \label{eq:HD_U}
~~~,
\end{align}
where the ``sil'' and ``car'' superscripts indicate silicate and carbonaceous grains, respectively. The frequency dependence of $I_\nu$ and $P_\nu$ for each grain type is precomputed in the context of the \citet{Hensley+Draine_2017c} grain model. The relative abundance of silicate and carbonaceous grains is fixed to the default employed by \citet{Hensley+Draine_2017c} and normalized to yield the desired 353\,GHz total intensity of 50\,$\mu$K$_{\rm RJ}$. We note that in this model the carbonaceous grains are unaligned and therefore do not produce polarized emission (i.e., $Q_\nu^{\rm car} = U_\nu^{\rm car} = 0$). Therefore, $A_{d,1}^Q = A_{d,1}^U$ are adjusted to yield the desired 3.5\,$\mu$K polarized intensity at 353\,GHz. We refer to this as the {\it HD} model.

Finally, we consider a physical model in which the silicate grains have embedded iron inclusions that constitute 5\% of the grain volume. Denoting this grain type as ``sil+Fe,'' we have
\begin{align}
I_\nu^{\rm HD+Fe} &= A_{d,1}^I I_\nu^{\rm sil+Fe} + A_{d,2}^I I_\nu^{\rm car} \label{eq:HDFe_I} \\
Q_\nu^{\rm HD+Fe} &= A_{d,1}^Q Q_\nu^{\rm sil+Fe} + A_{d,2}^Q Q_\nu^{\rm car} \label{eq:HDFe_Q} \\
U_\nu^{\rm HD+Fe} &= A_{d,1}^U U_\nu^{\rm sil+Fe} + A_{d,2}^U U_\nu^{\rm car} \label{eq:HDFe_U}
~~~.
\end{align}
We note that the signs of $Q_\nu^{\rm sil+Fe}$ and $U_\nu^{\rm sil+Fe}$, and thus the polarization angle, depend on the relative importance of emission from the silicate and iron components. Therefore, this model has a frequency-dependent polarization fraction. We set the parameters of this model in an analogous way to the HD model, assuming the default relative abundances of silicate and carbonaceous grains, assuming that the carbonaceous grains produce unpolarized emission, and normalizing to the desired 50 and 5\,$\mu$K$_{\rm RJ}$ at 353\,GHz in intensity and polarization, respectively. We refer to this model as {\it HD+Fe}.

\subsection{Synchrotron}
\label{sec:sync}
As relativistic electrons spiral about the interstellar magnetic field, they emit photons at radio wavelengths known as synchrotron radiation. The spectrum of this radiation depends on the energy spectrum of the electrons. Over the frequency range of interest, the specific intensity is often approximated as a power law:
\begin{equation}
I_\nu^s = A_s^I \left(\frac{\nu}{\nu_0^s}\right)^{\beta_s}
~~~,
\end{equation}
where $A_s^I$ is the specific intensity at a reference frequency
$\nu_0^s$ and $\beta_s$ is the power law index.

This parameterization can be extended to polarization by assuming that
the polarized intensity spectrum has the same spectral shape as the
intensity spectrum:
\begin{equation}
P_\nu^s = A_s^P \left(\frac{\nu}{\nu_0^s}\right)^{\beta_s}
~~~,
\end{equation}
where $A_s^P$ is the specific polarized intensity at a reference frequency $\nu_0^s$, taken to be 30\,GHz. In this model, we assume the synchrotron polarization angle $\gamma_s$ is independent of frequency and is equal to the dust polarization angle at 353\,GHz.

To produce the adopted total and polarized synchrotron intensities, we adopt $A_s^I = 830$\,Jy\,sr$^{-1}$, $A_s^Q = A_s^U = 280$\,Jy\,sr$^{-1}$, and $\beta_s = -1.2$.

We note that the model of synchrotron presented here is a simplification, as it neglects known complexities such as curvature in the synchrotron SED and line of sight effects (cf. Section~\ref{sssec:mbb_cloud}). We defer investigation of the impact of these complexities to future work, as we focus the present analysis principally on dust emission.

\subsection{Anomalous Microwave Emission}
\label{ssec:ame}
Interaction with gas atoms in the interstellar medium causes dust grains to rotate, and grains possessing an electric dipole moment radiate as they rotate \citep{Draine+Lazarian_1998b}. Ultrasmall grains (radius $a \lesssim 10\,$\AA) have rotational frequencies of order 30\,GHz and thus contribute to Galactic microwave emission. The spinning dust mechanism is believed to be responsible for the ``anomalous microwave emission'' (AME), a prominent dust-correlated component of the Galactic microwave emission between $\sim10-40$\,GHz.

\begin{figure*}
	\includegraphics[width=\textwidth]{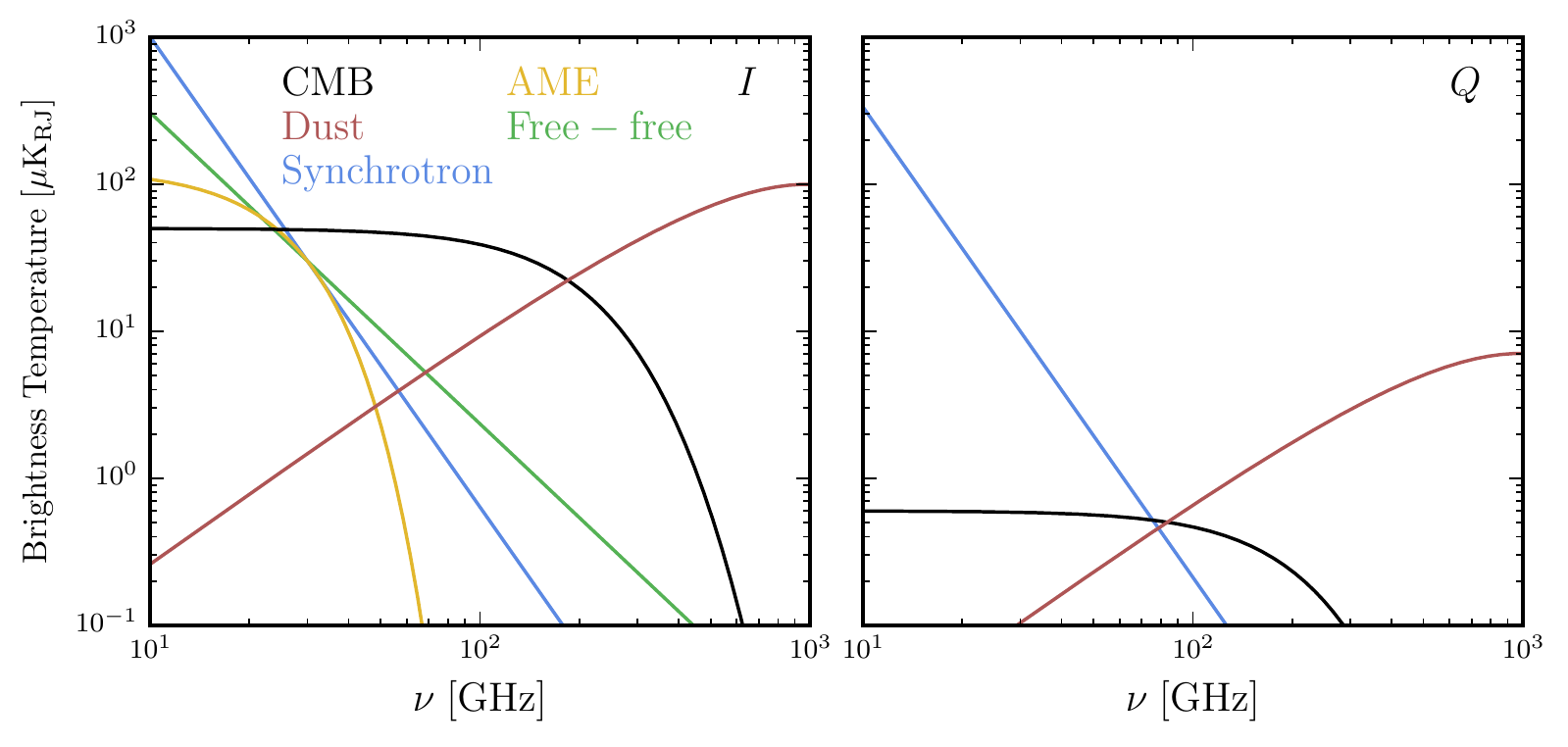}
	\caption{Input SEDs for the CMB, synchrotron, dust (MBB model), AME, and free-free in Stokes $I$ and $Q$. Stokes $U$, not shown, is identical to Stokes $Q$ for all components. AME and free-free are assumed to be unpolarized.}
	\label{fig:other_fgs}
\end{figure*}

While the AME spectrum has been measured in detail in some Galactic clouds \citep{Planck_Int_XV}, the SED of the diffuse Galactic AME appears to peak at lower frequencies and is much more poorly constrained \citep{MivilleDeschenes+etal_2008,Planck_2015_X}. Further, cloud-to-cloud variations suggest large variability in the AME SED, with peak frequencies typically ranging from 20-35\,GHz, though as high as $\simeq$50\,GHz \citep{Planck_Int_XV}. Theoretical constraints on the spinning dust SED are also weak due to sensitivity of the SED to conditions in the ambient interstellar medium, with different AME SEDs predicted for, e.g., the warm and cold neutral media \citep{Draine+Lazarian_1998b, Ali-Haimoud+Hirata+Dickinson_2009}. The unknown grain size, charge, and dipole moment distributions lead to further uncertainties in the shape of the AME SED \citep{Hensley+Draine_2017b}.

In light of both the empirical and theoretical uncertainties, we seek a simple parameterization of the AME SED which can capture variability in the amplitude and peak frequency. For the purposes of this work, we adopt the functional form proposed by \citet{Draine+Hensley_2012} which includes a parameter governing the peak frequency $\nu_{\rm pk}$ of the spectrum:
\begin{equation}
I_\nu^{\rm AME} = A_{\rm AME}^I \left(\frac{\nu}{\nu_0^{\rm
  AME}}\right)^2\,{\rm exp}\left[1 - \left(\nu/\nu_{\rm pk}\right)^2\right]
~~~.
\end{equation}

The AME has yet to be detected definitively in polarization, though stringent upper limits ($p \lesssim 1$\%) have been placed in some specific regions \citep{Dickinson+Peel+Vidal_2011, Planck_2015_XXV,GenovaSantos+etal_2017} and for the large scale diffuse emission \citep{Kogut+etal_2007, Planck_Int_XXII}. Theoretical arguments suggest that energy quantization in ultrasmall grains dramatically suppresses the conversion of rotational to vibrational energy, preventing the grains from aligning with the interstellar magnetic field. If this mechanism is acting, the AME will be negligibly polarized \citep[$p \sim 10^{-8}$,][]{Draine+Hensley_2016}. We therefore adopt an unpolarized AME component for the purposes of this work on both empirical and theoretical grounds.

To produce the adopted total 30\,GHz AME intensity of 30\,$\mu$K$_{\rm RJ}$, we take $\nu_{\rm pk} = 25$\,GHz and $A_{\rm AME}^I = 1300$\,Jy\,sr$^{-1}$.

\subsection{Free-Free}
\label{ssec:ff}
Free-free emission arises from the acceleration of electrons due to the electric fields of ions. The free-free spectrum is well-known both empirically and theoretically, being well-approximated by a simple power law in the optically thin limit:

\begin{equation}
I_\nu^{\rm ff} = A_{\rm ff}^I \left(\frac{\nu}{\nu_0^{\rm
      ff}}\right)^{-0.12}
~~~,
\end{equation}
where $A_{\rm ff}^I$ is the amplitude at a reference frequency $\nu_0^{\rm ff}$, taken to be 30\,GHz.

Free-free emission is inherently unpolarized due to its isotropic nature. However, near the edges of H{\sc ii} regions, Thomson scattering can induce a low level of polarization, estimated to be at the $\sim 10$\% level when observing in the Galactic plane at high resolution \citep{Keating+etal_1998,Macellari+etal_2011}. At the relatively low resolutions analyzed here, the polarization is expected to be much smaller. Indeed, empirically, the free-free polarization appears to be less than 1\% \citep{Macellari+etal_2011,Planck_2015_XXV}. Thus, for the purposes of this work, we neglect a potential free-free contribution to the total polarized emission.

To produce the adopted total 30\,GHz free-free intensity of 30\,$\mu$K$_{\rm RJ}$, we take $A_{\rm ff}^I = 830$\,Jy\,sr$^{-1}$.

\subsection{Other Emission Components}
The foregoing sections, while extensive, have not been exhaustive of the numerous emission mechanisms in the frequency range of interest. Line emission (notably CO), the cosmic infrared background (CIB), and zodiacal light will all contribute to the sky signal observed by a real experiment. As these components are subdominant in intensity and are expected to be negligibly polarized, we neglect them for the purposes of this analysis.

\subsection{Summary}

We illustrate the SEDs of the emission components employed in this work in Figure~\ref{fig:other_fgs}. In our analysis, all components are kept fixed to these SEDs with the exception of the dust emission. The various dust model SEDs are summarized in Figure~\ref{fig:dust_models}.

\section{Component Separation Simulations}
\label{sec:method}

To understand how different types of dust contamination can affect the recovery of CMB temperature and polarization maps, we performed a large number of simulations of single sky pixels across multiple bands. The simulations covered a wide range of assumptions about dust physics and instrumental design, and included multiple ($\sim 100$) noise realizations per dust model and band configuration, so that the statistical properties of the component separation procedure could be studied. In this paper, we consider a single foreground removal algorithm that uses a Bayesian model fitting procedure on multi-band, single-pixel data. Single-pixel foreground fitting is one of the most conservative foreground removal strategies, as it does not require any assumptions to be made about the spatial distribution of the foreground components. The need to constrain multiple foreground degrees of freedom per pixel is also an important driver of the many-band designs of several future CMB experiments.

In this section, we describe the details of the single-pixel simulations and the MCMC model fitting procedure that was subsequently applied to them.

\subsection{Single-pixel simulations}
\label{sec:spsims}

For each dust model, we simulated 100 single-pixel data vectors with different noise realizations over a grid of minimum and maximum frequencies. A 7-band instrumental design was assumed in all cases, with minimum and maximum frequencies of $\nu_{\rm min} = (20, 30, 40)$ GHz and $\nu_{\rm max} = (300, 400, 500, 600, 700, 800)$ GHz respectively. The bands were spread across the frequency range with a constant logarithmic interval between them. No bandpass integration was performed, so we have essentially assumed a delta-function bandpass at all frequencies.

\begin{figure}[t]
    \centering
    \includegraphics[width=0.5\textwidth]{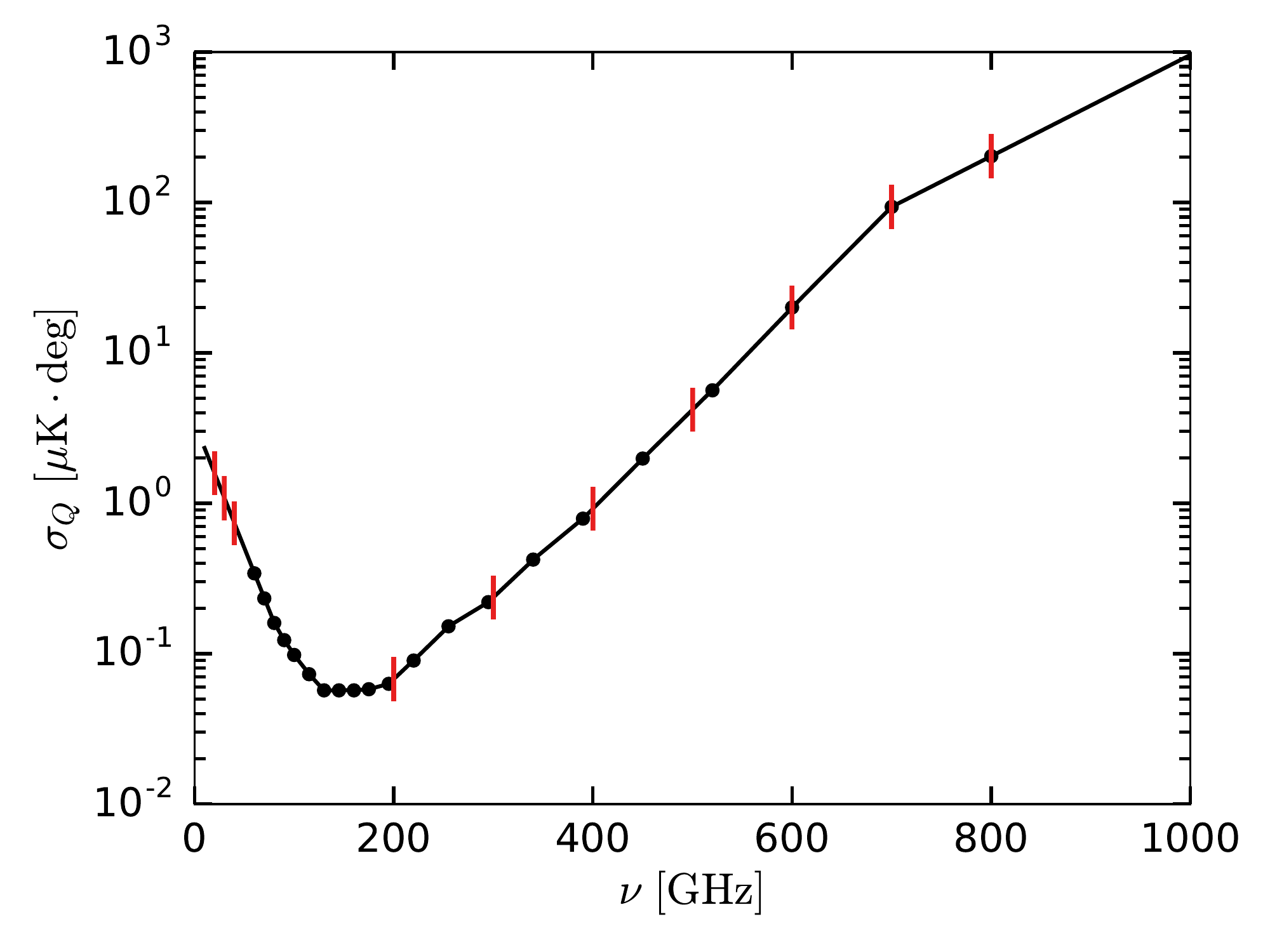}
    \caption{Assumed Stokes Q noise rms as a function of frequency for a future CMB polarization experiment. The U noise rms is identical, and the I noise rms is smaller by a factor of $\sqrt{2}$. We base the noise curve on the COrE+ (Extended) specifications from \citet{Remazeilles+etal_2016} (black points), with a log-linear extrapolation to higher and lower frequencies. The minimum/maximum frequencies that were used are marked on the curve as red vertical lines.}
    \label{fig:noise_models} 
\end{figure}

The noise at each frequency was obtained by interpolating the CoRE+ Stokes Q noise curve from \cite{Remazeilles+etal_2016}, shown in Figure~\ref{fig:noise_models}. We assumed a log-linear extension of the noise curve at frequencies lower than 60\,GHz, and a circular, 1$^\circ$ FWHM beam in each frequency channel. This angular resolution should be sufficient for typical B-mode analyses that target the low-$\ell$ primordial B-mode signal, although higher resolution experiments have the advantage of being able to estimate the CMB lensing contribution more effectively \citep{2012JCAP...06..014S}. The Q and U polarization channels were assumed to have equal noise levels, with the Stokes I noise rms lower by a factor of $\sqrt{2}$. The noise was assumed Gaussian and uncorrelated between bands. These noise levels are typical of proposed space-based CMB polarization experiments and so provide a benchmark for our analysis.  Joint optimization of frequency coverage and the signal to noise in each frequency band is beyond the scope of this study, but we note that the details of the results presented here would change with the adoption of a different noise curve.

To enable a fair comparison, the model for each component of the sky signal was kept the same between noise realizations and choices of minimum/maximum frequency. The assumptions for each of the non-dust components are summarized in Table~\ref{tbl:components}; I, Q, and U amplitudes were chosen to be representative of the recovered amplitudes from the Planck 2015 Commander-Ruler marginal foreground maps \citep{Planck_2015_X}, smoothed to 1$^\circ$, and with a sky cut of $|b| \ge 30^\circ$, as discussed in Section~\ref{sec:fg_amp}. Spectral parameters were chosen to be broadly consistent with current constraints from the literature. While most of these parameters are expected to vary across the sky (some of them significantly), the chosen values should be representative of a `typical' pixel after excluding the Galactic plane.

Note that we assumed the CMB contribution to be free of lensing contamination, which we would not be able to estimate from a single pixel anyway, and that polarization leakage and other instrumental effects can be ignored.

For the thermal dust component, seven different models were studied, as described in Section~\ref{sec:thdust} and summarized in Table~\ref{table:dust_models}. The total intensity and polarization amplitudes at 353\,GHz were chosen to be the same across all models. The adopted spectral parameters are motivated by the physical effects under study and by models in the literature, as detailed in Section~\ref{sec:thdust}.

\subsection{Single-pixel foreground fitting procedure}
\label{sec:spfits}

With the simulated data in hand, we applied a Bayesian foreground model fitting procedure to each data vector to recover the CMB I, Q, and U amplitudes. This uses a Markov Chain Monte Carlo (MCMC) method to sample from the joint posterior of the amplitude and spectral parameters of all of the component models. The CMB amplitudes can then be obtained by marginalizing over all the other parameters.

\begin{table}[t]
 \centering
 \begin{tabular}{lll}
 \hline
 MBB & Min. & Max. \\
 \hline
 $T_d$ & 16 & 24 \\
 $\beta_d$ & 1.4 & 1.8 \\
 \hline
 2MBB & Min. & Max. \\
 \hline
 $T_{d,1}$, $T_{d,2}$ & 5 & 30 \\
 $\beta_d$ & 1.1 & 1.8 \\
 $\Delta \beta_d$ & $-1.8$ & 1.8 \\
 $f_I$ & 0 & 1 \\
 $f_Q = f_U$ & $-2$ & 2 \\
 \hline
 Other & Min. & Max. \\
 \hline
 $\beta_s$ & $-1.6$ & $-0.8$ \\
 $\nu_{\rm peak}^{\rm AME}$ & 15 & 35 \\
 $A^I_X$ & 0 & $\infty$\\
 \hline
 \end{tabular}
 
 \caption{Prior ranges of various model parameters. Temperatures are in K and frequencies in GHz; all other spectral parameters are dimensionless. Parameters not included in this table were assumed to lie in the range $[-\infty, \infty]$. Note that we set $f_Q = f_U$ to reduce the number of free parameters in the 2MBB model.}
 \label{tbl:priors}
\end{table}

\begin{figure*}[t]
    \hspace{-1em}
	\includegraphics[width=1.03\textwidth]{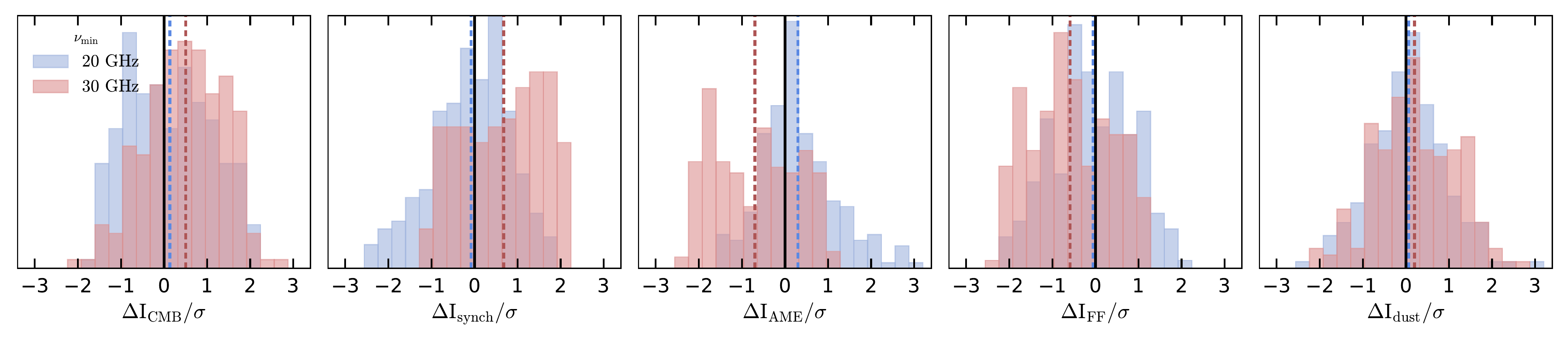}
	\caption{Bias in the maximum likelihood estimates of the Stokes I amplitudes of 5 components, for a fit to temperature + polarization data containing an MBB dust component, plus CMB, AME, free-free, and synchrotron. Statistics are shown for 200 noise realizations, with $\nu_{\rm max} = 500$ GHz, and two different values of $\nu_{\rm min}$. The dashed lines show the mean bias over all 200 realizations. If $\nu_{\rm min}$ is too high, degeneracies between the low-frequency components result in biased amplitude estimates.}
	\label{fig:bias_tp}
\end{figure*}

The sampling procedure is set up as follows. We first construct a Gaussian likelihood, of the form
\begin{equation} \label{eq:loglike}
\log \mathcal{L}(\vec{\theta}) \sim -\frac{1}{2} \left (\vec{d} - \vec{s}(\vec{\theta}) \right )^{\rm T} N^{-1} \left (\vec{d} - \vec{s}(\vec{\theta}) \right) ,
\end{equation}
where the data vector $\vec{d}$ consists of temperature values in each frequency band and polarization. The combined signal vector, $\vec{s}$, is constructed by summing the contributions from each component model in each frequency band and polarization for a given set of parameters, $\vec{\theta}$. The noise covariance, $N$, is assumed diagonal, and is identical to the noise covariance that was used when generating the simulations.

We are interested in the effects of assuming the wrong form for the dust model, i.e., what happens when we fit a phenomenological model to one of the more complicated physically-motivated dust models described in Section~\ref{sec:thdust}. For each of the input models described above, we fit two models to the simulated data: the simple modified blackbody (MBB) model, and the generalized two-component MBB (2MBB) model in which both components are assumed to have the same polarization angle. The MBB model has often been used in analyses of CMB data, and only has two free spectral parameters, $\beta_d$ and $T_d$. The 2MBB model is expected to better encapsulate complex dust physics, at the cost of introducing several extra parameters. For all other foreground components, we fit the same model that was used to generate the simulated data vector, with all relevant parameters being included in the sampling process.

For each simulated data vector and choice of dust model to be fit, we used the {\tt emcee} affine-invariant ensemble sampler \citep{Foreman-Mackey+etal_2013} to return samples from the joint posterior, using the likelihood from Eq.~\ref{eq:loglike}. All parameters were taken to have uniform priors over relatively broad ranges, given in Table~\ref{tbl:priors}. The sampler was run for 10,000 steps from each of 100 walkers, with a burn-in of 8,000 steps being discarded from each. This burn-in period was determined to be sufficient for both the MBB and 2MBB fits, following a series of convergence studies that checked sensitivity of the means and standard deviations of the CMB amplitudes to burn-in length. Long burn-in periods were needed to allow the walkers time to find the maximum likelihood region on what were sometimes quite complex likelihood surfaces that exhibited multiple degeneracies. Note that, due to the long correlation times in cases where there were strong degeneracies, some of the chains may not have formally converged. Further convergence tests suggest that this is not a serious problem, as only small shifts in the posterior means (a few percent of a standard deviation) were observed when running much longer chains with 100,000 samples per walker. This is subdominant to the error on the posterior mean due to the finite number of noise realizations that were used. Also note that we do not thin the chains, relying instead on combining the samples from the 100 walkers to reduce sample correlation effects.

Each worker was started from an initial position with a small random displacement from the input parameter values used in the simulation. This was done to ensure that the `correct' CMB and foreground parameters would at least be explored in case of multi-modal distributions or parameter degeneracies. The exception was the parameters of the dust model used in the fits, which were set to the same values no matter the input model (but which at least started with the correct amplitude at 353\,GHz).

After running and processing the MCMC chains, we calculated a set of summary statistics from the CMB amplitude parameter chains, i.e., the recovered CMB I, Q, and U amplitude pdfs marginalized over all other parameters. This was done for each of the 100 noise realizations for each pair of input and fitting dust models. The distributions of these summary statistics are analyzed in detail in the next two sections. From the MCMC chain for each noise realization, we kept the mean and standard deviation of each parameter, as well as its value at the maximum a posteriori probability (MAP), which is coincident with the point of minimum $\chi^2$. These were used to calculate the (error-normalized) bias from the true (input) value of each parameter,
\begin{equation}
\frac{\Delta \theta}{\sigma_\theta} \equiv \frac{\bar{\theta} - \theta_{\rm true}}{\sigma_\theta},
\end{equation}
where $\theta$ is some parameter, the bar denotes either the mean or MAP estimate from the MCMC chain, and $\sigma_\theta$ is the marginal standard deviation of that parameter estimated from the MCMC chain. In the rest of the paper, we study the distributions of these summary statistics over the 100 noise realizations. Summary statistics that run over noise realizations are denoted by angle brackets, for example $\langle \Delta \theta \rangle$ denotes the mean of the bias over the 100 noise realizations.

\section{Joint Temperature and Polarization Analysis}
\label{sec:temp_pol}

In this section, we perform an example analysis that uses both temperature (Stokes I) and polarization (Stokes Q and U) data to constrain the CMB and foreground models. Naively, one would expect the inclusion of an I channel to improve the CMB polarization constraints, as it should help to constrain parameters of the foreground models that are common to I and Q/U. As we will see here however, this is not necessarily the case due to the complexity of the low-frequency temperature foregrounds.

\begin{figure}
	\hspace{-1em}
	\includegraphics[width=0.5\textwidth]{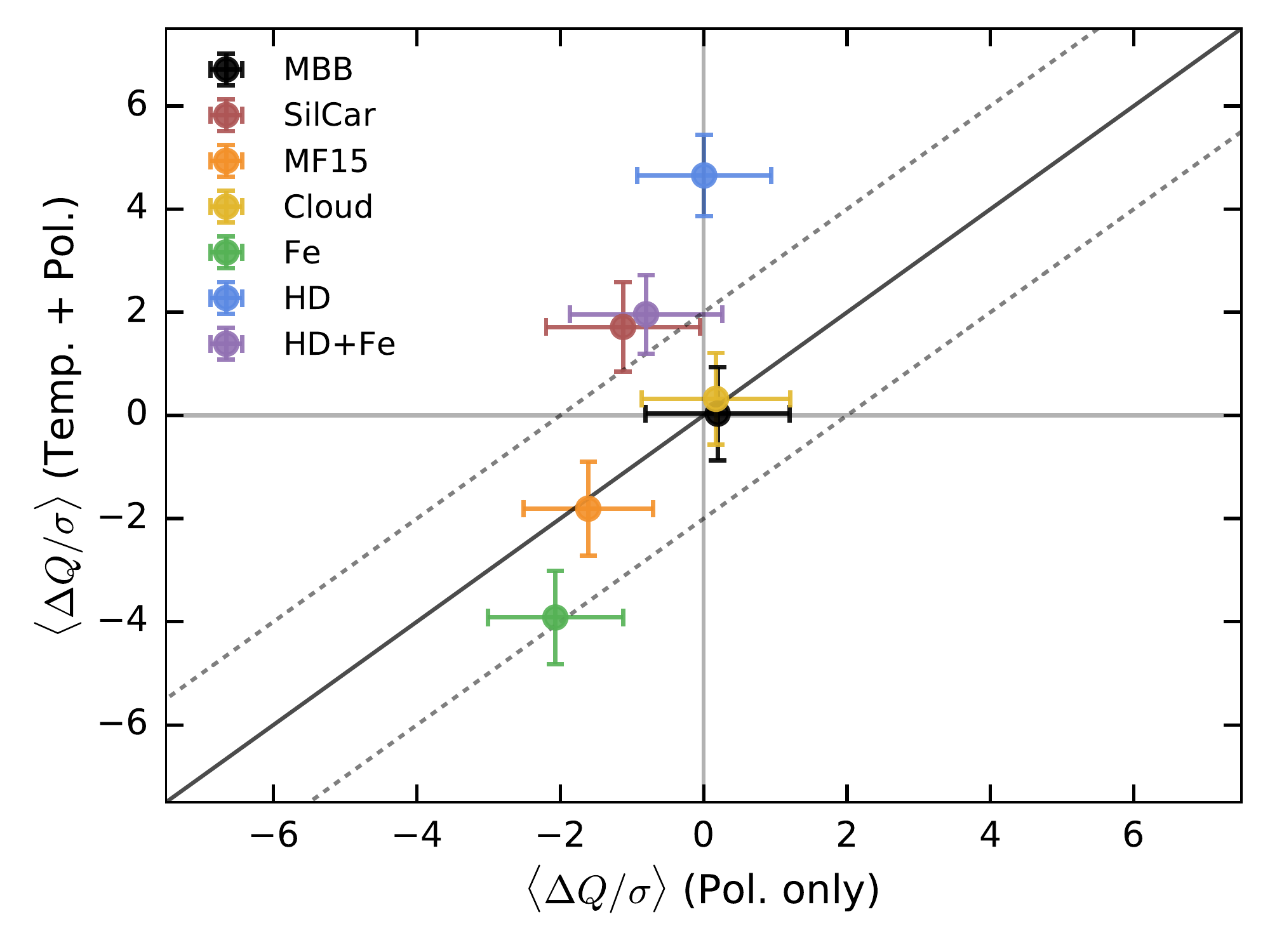}
	\caption{Comparison of the mean bias of the recovered CMB Q amplitude between temperature + polarization and polarization-only analyses. The points shown are for fits with an MBB dust model, and a band configuration with $\nu_{\rm min}, \nu_{\rm max} = 20,\, 500$ GHz. The errorbars show the standard deviation of the bias, while the solid diagonal line marks where the two analyses would have equal bias. Most models show a tendency towards a greater degree of bias when temperature information is included.}
	\label{fig:tp_vs_p}
\end{figure}

In Figure~\ref{fig:bias_tp}, we show the result of fitting CMB and four foreground components (synchrotron, free-free, AME, and dust) to temperature plus polarization data over seven frequency bands for two different choices of $\nu_{\rm min}$, up to a $\nu_{\rm max}$ of 500\,GHz. In this example, we used a simple MBB model as both the input dust model and the model for the fitting procedure. Since the models to be fitted are the same as the input models, we expect to be able to recover the input model amplitudes and parameters without bias. Figure~\ref{fig:bias_tp} shows that this is more or less the case for a band configuration with $\nu_{\rm min} = 20$ GHz, but that there is a significant bias for several components when $\nu_{\rm min} = 30$ GHz.

This is caused by a degeneracy between several of the low-frequency foreground components, which have similar spectral behaviors around $\sim 30$ GHz (see Figure~\ref{fig:other_fgs}). Without at least one band below $30$\,GHz, there is insufficient information to reliably distinguish between synchrotron, free-free, and AME. Incorrect inferences about these components can then lead to them being under- or over-subtracted at higher frequencies, leaving residuals that systematically bias the CMB. In the cases shown in Figure~\ref{fig:bias_tp}, the bias on the CMB Q and U amplitudes is relatively small ($\sim 0.2 \sigma$), but this is just the simplest case. For more complex dust models, the bias can be more substantial. Figure~\ref{fig:tp_vs_p} compares the bias on the CMB Q amplitude for all seven dust models, for analyses that use temperature and polarization (T+P) information, versus polarization information alone. In four out of the seven cases, the bias is significantly larger for the T+P analysis -- and this is for the relatively optimistic scenario where $\nu_{\rm min} = 20$ GHz. Note that such biases would likely be identifiable however, as the degeneracies would be apparent in the MCMC chains; the bias shown in Figure~\ref{fig:bias_tp} is based on a summary statistic, which obscures the existence of degeneracies.

We draw two conclusions from this. Firstly, an experiment using joint temperature plus polarization foreground fits will likely require frequency coverage below 30\,GHz to help break the low-frequency degeneracy. In the absence of low-frequency bands, one can use an amalgamated low-frequency foreground component instead of separate physical synchrotron, free-free, and AME models, as was done in \cite{Planck_2013_XII}. This is not an unreasonable approach, but runs the risk of introducing subtle model errors, especially if information from the `combined' low-frequency foreground in temperature is used to make inferences about its polarization properties.

Secondly, if the aim of an experiment is to recover the polarized CMB as accurately as possible, then the value of the additional foreground information provided by the I channel is likely outweighed by the increased complexity of the low-frequency temperature foregrounds. This is especially the case if one considers that we have used quite simplistic models for the synchrotron and AME components here, ignoring potential complications such as spectral curvature, shape of the AME SED, etc. 

Additionally, we have assumed the optimistic case that the spectral parameters of the synchrotron and dust components are the same in both temperature and polarization. However, the {\it Planck} observations indicate that the dust spectral index $\beta_d$ is systematically different in temperature and polarization \citep{Planck_Int_XXII}. To account for this, one could decouple the spectral parameters used in the I and Q/U channels for these components, but then the I channel provides no extra spectral information about the polarized foregrounds. As such, an analysis that uses only polarization information, without trying to model the temperature components, is likely to be simpler and more robust. This is the strategy that we pursue throughout the rest of the paper.

\section{Polarization-Only Analysis}
\label{sec:results}

In the following sections, we describe the results of fitting the MBB and 2MBB models to the seven dust models described in Section~\ref{sec:thdust}, using polarization information alone. In particular, we focus on the bias that is induced in the recovered CMB Q and U amplitudes due to fitting an `incorrect' dust model to the data, and whether the unsuitability of the fitting model can be identified by inspecting the $\chi^2$ goodness of fit statistic. While large biases are unwelcome in any situation, it is most problematic if one cannot identify that the results are probably biased and thus take some corrective action (like re-fitting the data with a more complex dust model). As such, models that are strongly biased but still yield low $\chi^2$ values are the most dangerous.

Our main results for the bias and $\chi^2$ are shown in Figs.~\ref{fig:chi2_mbb} and \ref{fig:bias_mbb}. Numerical values of the bias, $\chi^2$, and CMB signal-to-noise are given for a few example band configurations in Table~\ref{tbl:stats}.

\subsection{Modified blackbody models}
\label{sec:fit_mbb}

\begin{figure*}
    \hspace{-1em}
	\includegraphics[width=0.49\textwidth]{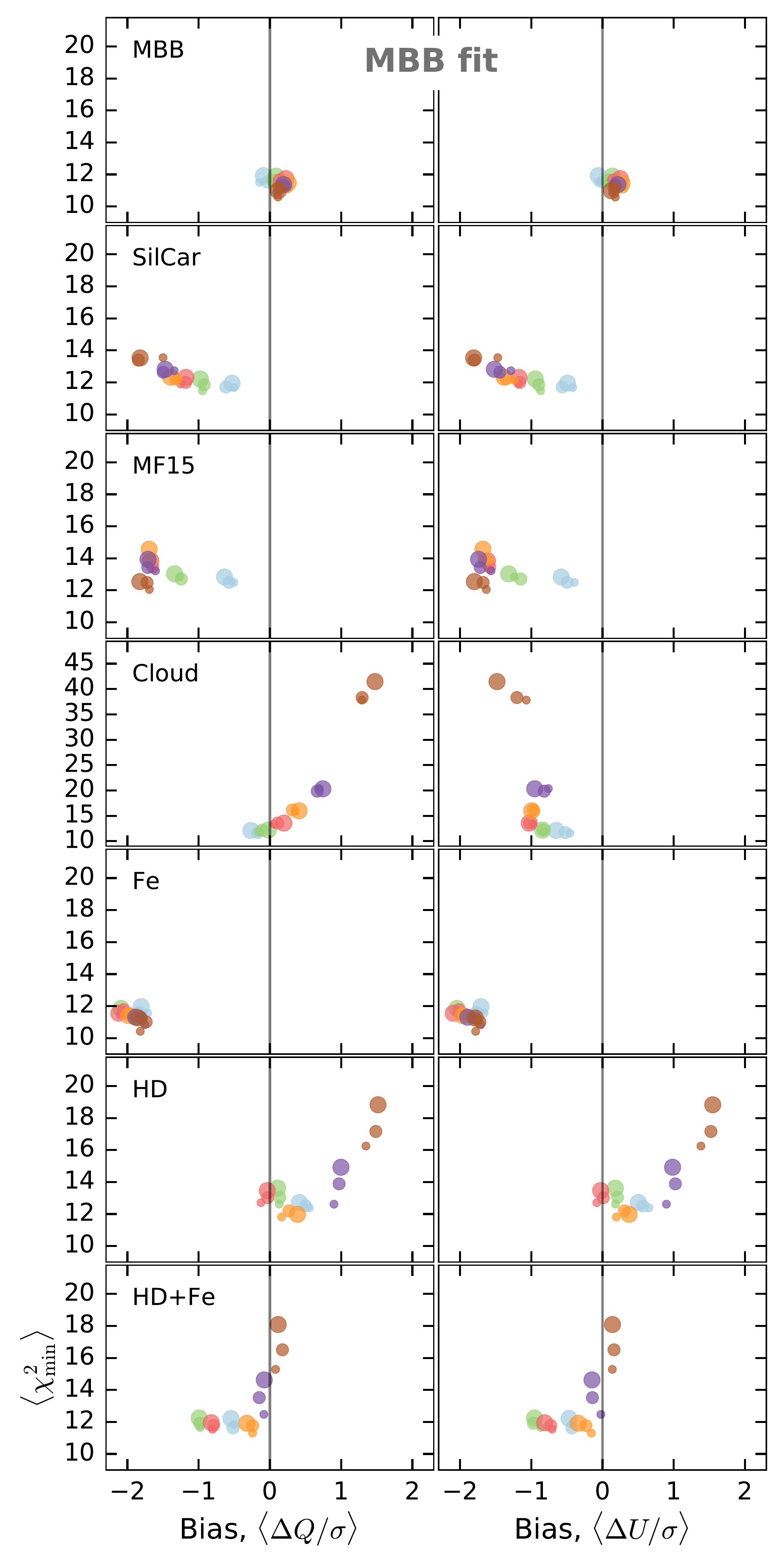}
    \includegraphics[width=0.49\textwidth]{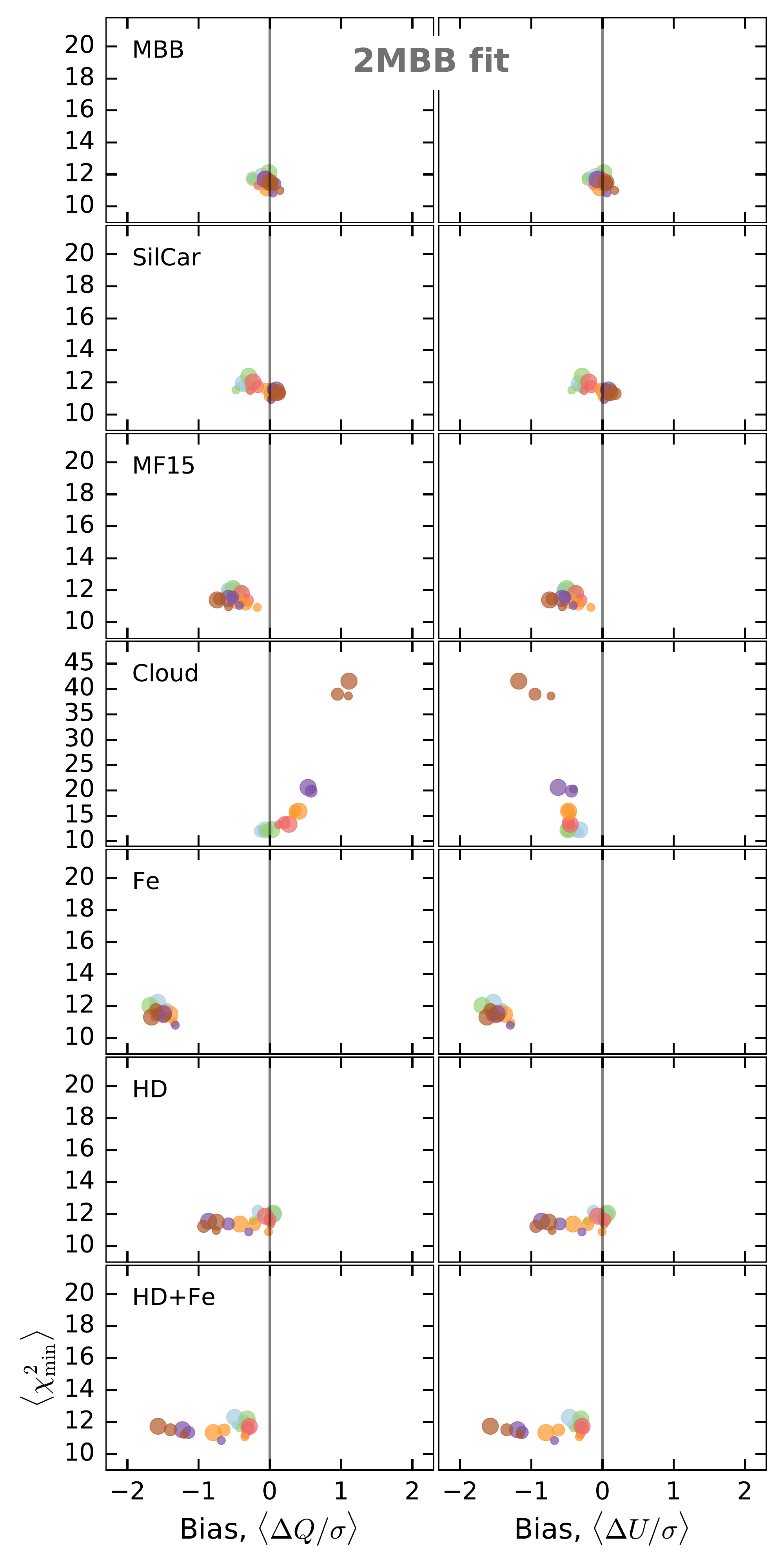}
	\caption{Mean bias of the Stokes Q and U CMB amplitudes (x axis) vs. the median of the minimum $\chi^2$ (y axis) over 100 noise realizations for simulations with several different dust models. The size and color of the points denote different choices of $\nu_{\rm min}$ and $\nu_{\rm max}$ respectively (see key below), while the left two and right two panels show results for when MBB and 2MBB models are used in the fits respectively. All fits were performed by fitting CMB + synchrotron + dust components to polarization data only. The biases are calculated using the maximum a posteriori probability (minimum $\chi^2$) values for the CMB amplitudes for each noise realization.}
	\label{fig:chi2_mbb}
\end{figure*}

\begin{figure*}
    \centering
	\includegraphics[width=0.45\textwidth]{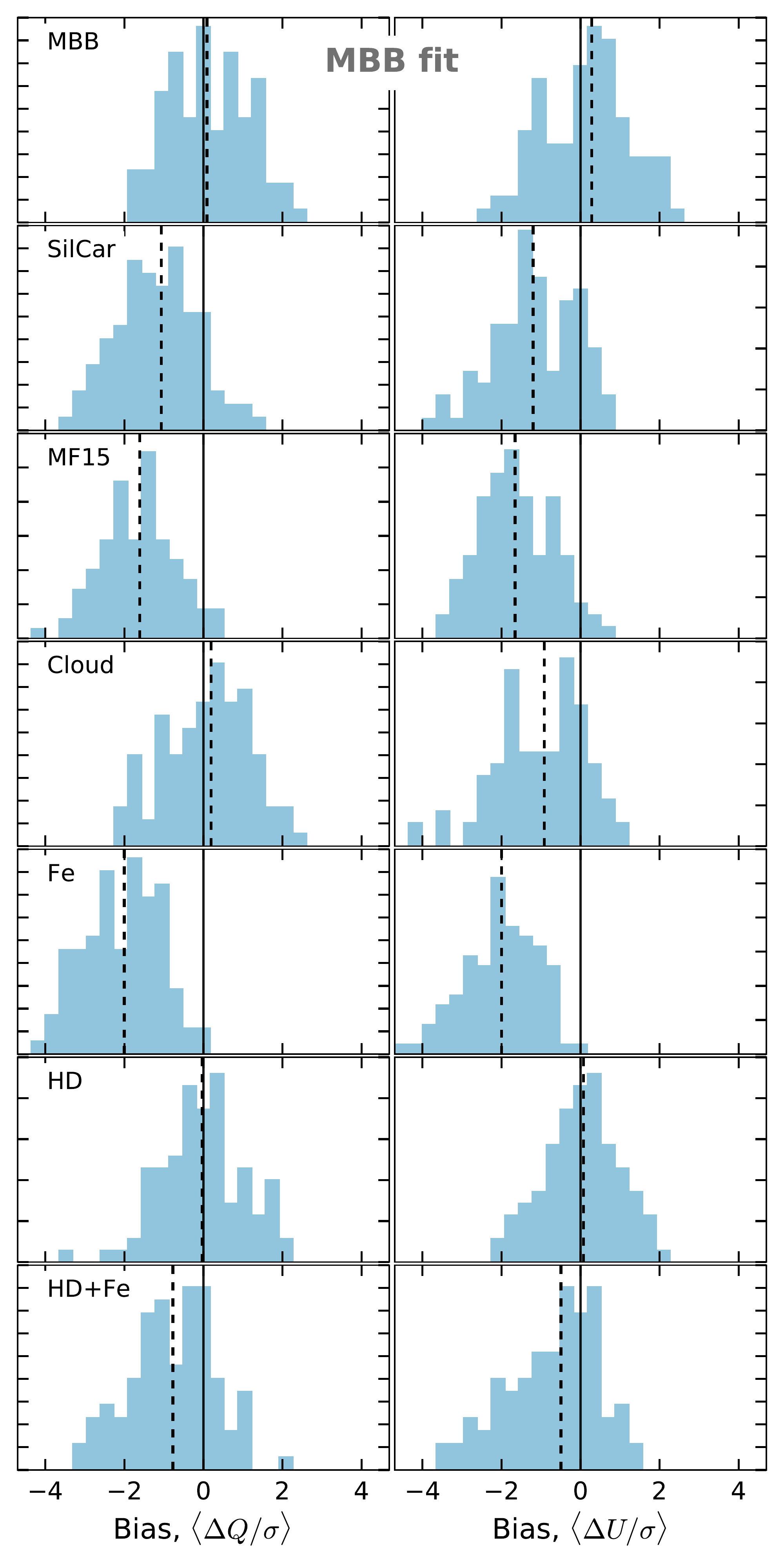}
    \includegraphics[width=0.45\textwidth]{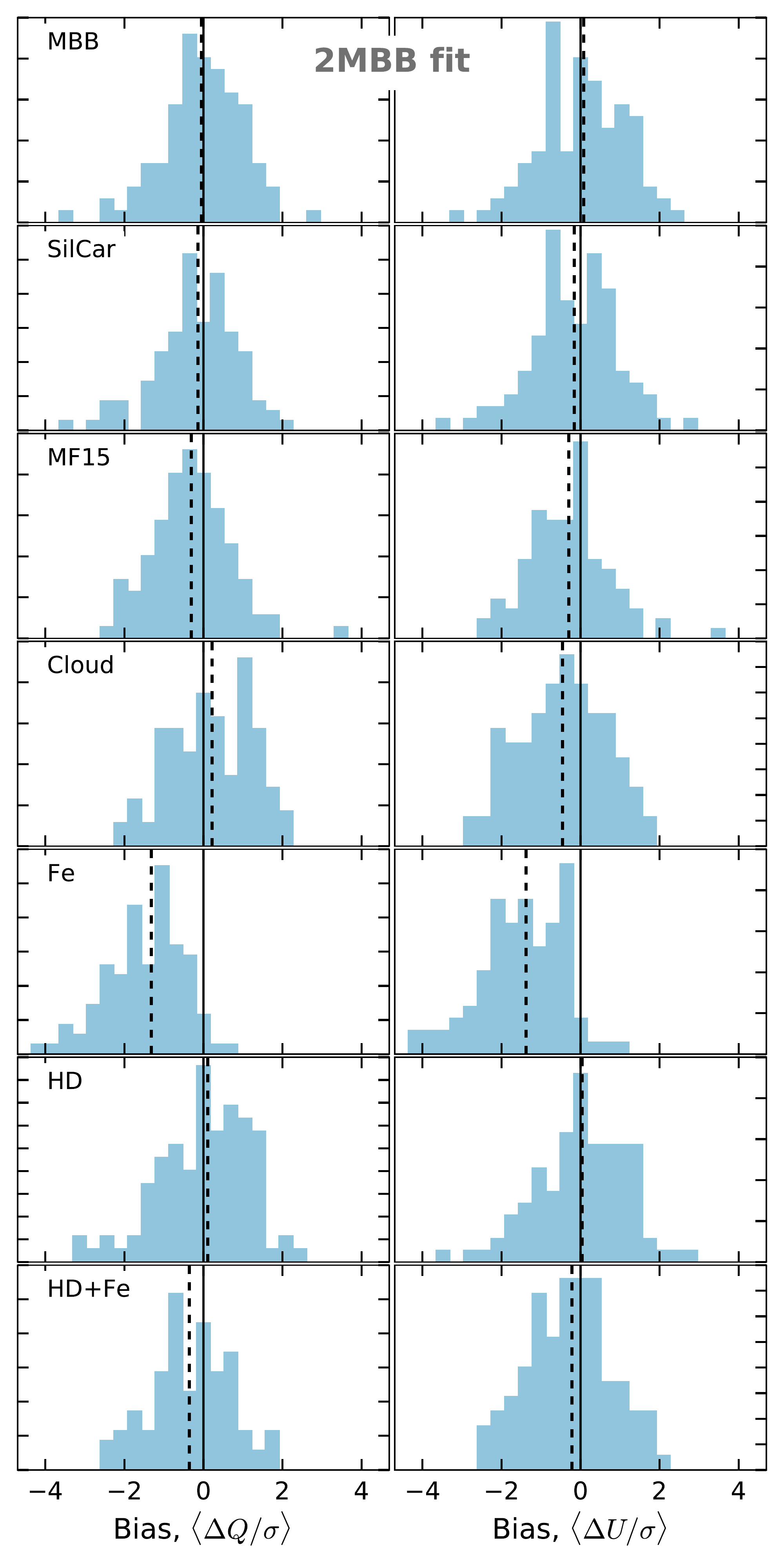}
	\caption{Bias in Stokes Q and U CMB amplitudes over 100 noise realizations, for the same dust models and fitting procedure as in Figure~\ref{fig:chi2_mbb}. The left two and right two panels again show the results for when MBB and 2MBB models are used to fit the dust component. The dashed vertical lines show the median bias. These results are shown for $\nu_{\rm min}, \nu_{\rm max} = 30, 500$ GHz.}
	\label{fig:bias_mbb}
\end{figure*}

We begin by fitting the MBB and 2MBB models to simulations that use the simple MBB model described in Section~\ref{sec:mbb} as the input dust component. As we illustrate in the top panels of Figs.~\ref{fig:chi2_mbb} and \ref{fig:bias_mbb}, both models recover the input CMB with minimal bias, as expected, with little dependence on the band configuration. The analysis techniques and summary statistics employed therefore appear unbiased and robust. We note however that the extra parameters introduced in the 2MBB model come at the expense of reduced signal to noise on the recovered CMB amplitudes (see Table~\ref{tbl:stats}).

Having performed these consistency tests, we now turn to the two 2MBB models discussed in Section~\ref{sec:2mbb}: the model with silicate and carbonaceous grains (SilCar) and the one based on the empirical fits of \citet{Meisner+Finkbeiner_2015} (MF15). When fitting either model with the MBB fitting function, the recovered CMB amplitudes are biased by 1--2$\sigma$, depending on the band configuration. 

\begin{figure}[t]
	\includegraphics[width=0.48\textwidth]{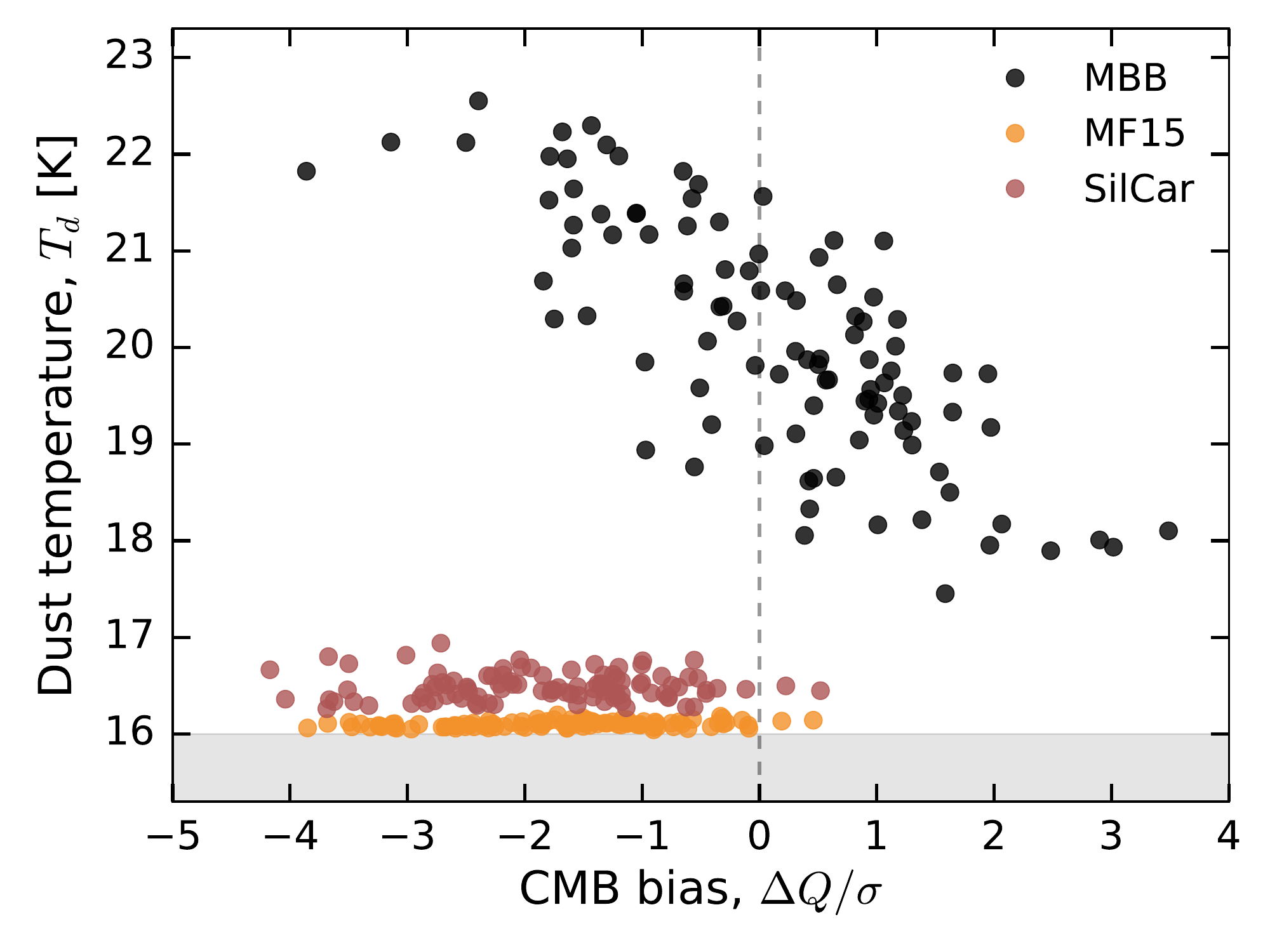}
	\caption{The bias in the maximum a posteriori CMB Q amplitude and dust temperature for MBB fits to polarization only data, with $\nu_{\rm min}, \nu_{\rm max} = 30, 800$ GHz. The gray band shows the region excluded by a prior on $T_d$. Each point shows the result for one of the 100 noise realizations.}
	\label{fig:dustT_mbb}
\end{figure}

In principle, the SilCar model should be perfectly described by the MBB fitting function in polarization only, as the carbonaceous grain component is unpolarized, leaving only a single polarized MBB component associated with the silicate grains. Nevertheless, the fits are biased. This is due to our choice of dust temperature prior. Guided by previous studies fitting the FIR dust emission, we naively selected a uniform prior on the dust temperature of $T_d \in [16, 24] \,{\rm K}$. This is a considerably broader range than was used in the \cite{Planck_2015_X} analysis for example, which assumed a Gaussian prior on $T_d$ with mean 23\,K and a standard deviation of 3\,K. Our prior is appropriate when fitting the SED in total intensity, but if only the lower-temperature component is polarized, as is the case here \citep[since silicate grains are thought to run cooler than carbonaceous grains;][]{Li+Draine_2001}, then it is {\it not} appropriate for fitting the polarized SED. This is illustrated clearly in Figure~\ref{fig:dustT_mbb}, in which the maximum {\it a posteriori} dust temperatures for each noise realization are shown to cluster at the prior bound.

\begin{figure}[t]
    \centering
    \vspace{0.5em}
    \includegraphics[width=0.4\textwidth]{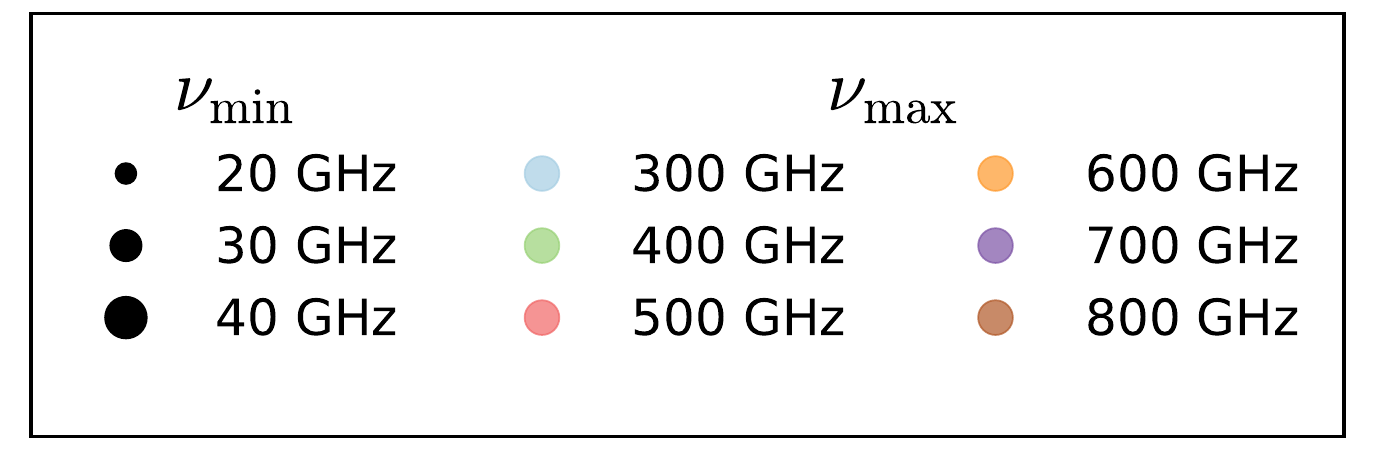}
    \vspace{-0.5em}
\end{figure}

A similar effect is observed when fitting MF15 with an MBB model. The dust component that dominates at low frequencies has a temperature of 9.75\,K, far below the prior bound, and so the same clustering effect occurs at the boundary (see Figure~\ref{fig:dustT_mbb}). The polarization spectrum is also complicated by the presence of a second component that dominates at high frequencies. The inadequacy of the MBB fit is seen in both the bias in the recovered CMB and, to a lesser degree, in slightly elevated $\chi^2$ values. The bias is reduced when high frequencies, where the complexities induced by the second dust component are most pronounced, are excluded from the analysis.

These problems would be identifiable from inspection of the MCMC chains, and so could be mitigated in a real analysis by expanding the priors. However, this does serve to highlight the risks of using temperature information to infer dust polarization properties. In particular, the dust polarization at CMB frequencies may be dominated by a dust component of significantly lower temperature than that responsible for most of the total intensity at frequencies near the dust peak. Therefore, dust temperature priors based on the total intensity data may be misleading. Further, at the noise levels expected in future experiments, the effects of the dust temperature on the {\it shape} of the polarized dust SED (i.e., departures from a pure Rayleigh-Jeans spectrum) are significant enough to impact the recovered CMB even at low frequencies. Joint fits to total and intensity and polarization must be done with care, and models employed in polarization only fits must be flexible enough to accommodate dust properties significantly different than what is observed in total intensity.

We now turn to fits of these models with a 2MBB model. As illustrated in Figs.~\ref{fig:chi2_mbb} and \ref{fig:bias_mbb}, the recovered CMB amplitudes are much less biased. This is expected, as the SilCar and MF15 models are both fully described by the 2MBB model (which also includes broader priors on its parameters). In the case of MF15, the fits are not completely unbiased. We discuss in more detail the biases that can arise when fitting with the 2MBB model in Section~\ref{sec:bias_sources}

Our conclusion from these results is that future CMB experiments will be sensitive to the additional complexity caused by multiple superposed dust components. Multi-component models of interstellar dust emission are well-motivated by empirical fits to the FIR dust SED \citep[e.g.][]{Finkbeiner+Davis+Schlegel_1999, Meisner+Finkbeiner_2015,Zheng+etal_2017} and by current dust models \citep[e.g.][]{Draine+Li_2007,Siebenmorgen+Voshchinnikov+Bagnulo_2014,Jones+etal_2017,Hensley+Draine_2017c}, and so forthcoming experiments should ensure that they can robustly remove multi-component dust contamination, particularly cases in which the components have significantly different properties.

\subsection{Models with line of sight effects}
\label{sec:depol}

Although the dust models discussed in Section~\ref{sec:fit_mbb} each featured multiple dust components, these components were all assumed to be aligned by the same magnetic field. In general, the Galactic magnetic field is not uniform along the line of sight, and polarized emission from dust aligned by a field of one direction will add in a vectorial way with emission from dust aligned by a field having a different direction. If the dust properties, such as composition or temperature, are also changing along the line of sight, then the frequency dependence of the resulting polarized emission can be complex and imperfectly correlated between frequencies. As demonstrated by \citet{Tassis+Pavlidou_2015} and \citet{Poh+Dodelson_2017}, failure to account for these line of sight effects can lead to biases in CMB fits.

To investigate this in greater detail, we employ the Cloud dust model described in Section~\ref{sssec:mbb_cloud}, in which the dust emission is assumed to arise from two clouds with different dust temperatures (15 and 20\,K) and magnetic field directions. As a result of the components having different polarization angles and SEDs, the ratio of $Q_\nu$ to $U_\nu$ varies with frequency.

The results of fitting the Cloud model with a simple MBB model are presented in Figs.~\ref{fig:chi2_mbb} and \ref{fig:bias_mbb}. While all frequency configurations are biased to a degree, those with the highest frequency bands suffer the highest bias. This is because the dust SEDs of the two clouds differ due to their temperature, and temperature effects are most pronounced at high frequencies. Thus, the more the frequency coverage extends to high frequencies, the less the simple model is able to account for the complexity of the emission. However, this is compensated by a dramatically increased $\chi^2$. Indeed, although models with lower $\nu_{\rm max}$ are less biased, they have more acceptable $\chi^2$ and thus pose a greater potential risk to the analysis.

\begin{figure}[t]
    \centering
    \vspace{-0.4em}
	\includegraphics[width=0.49\textwidth]{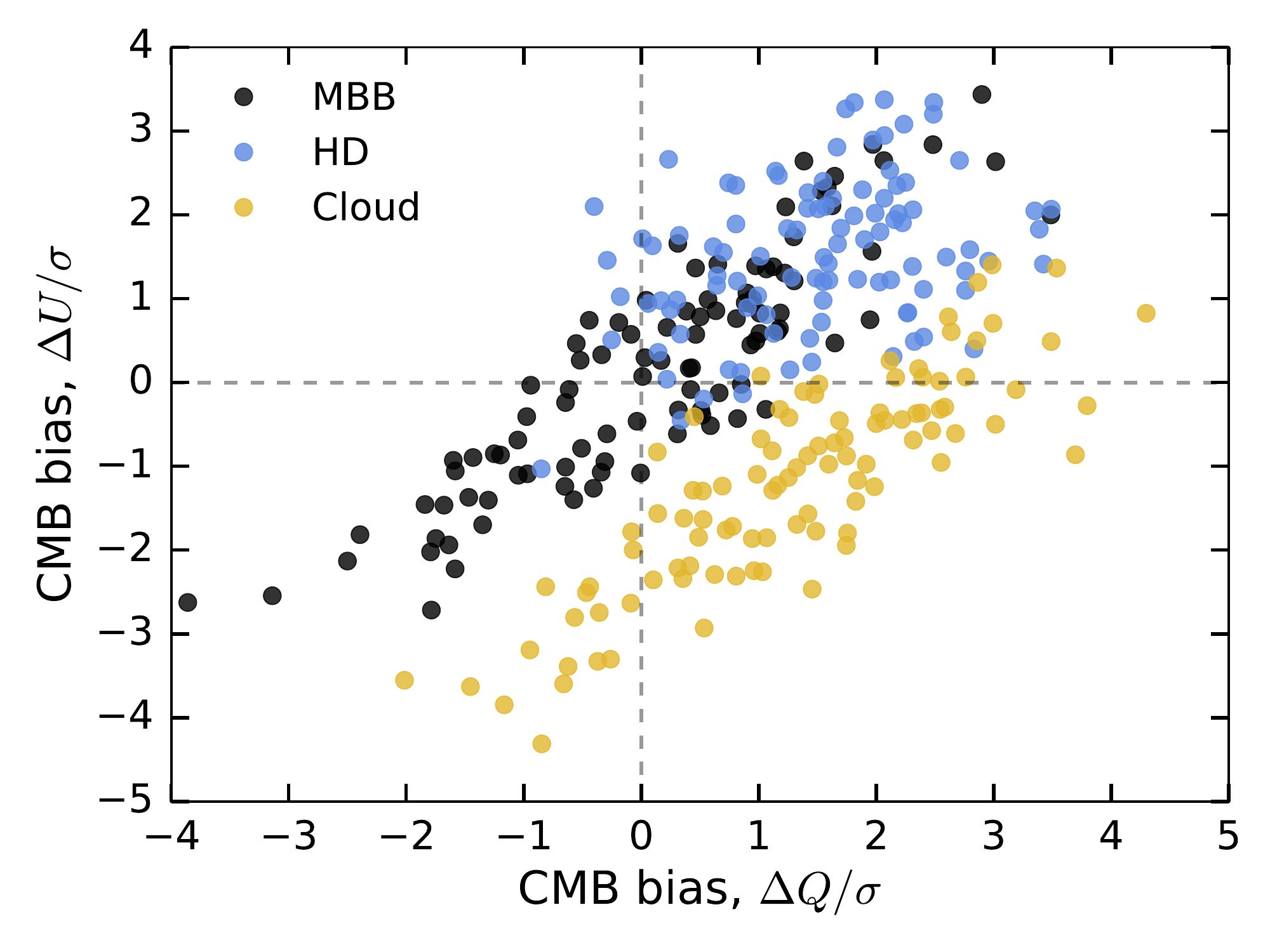}
	\caption{Bias of the CMB Q and U amplitudes for 100 noise realizations with three different input dust models. A simple MBB model was used to fit the data in all 3 cases. The results shown are for $\nu_{\rm min}, \nu_{\rm max} = 30, 800$ GHz, with polarization-only data.}
	\label{fig:bias_cmb_2d}
\end{figure}

The effect of the frequency-dependent polarization angle is illustrated in another way in Figure~\ref{fig:bias_cmb_2d}, which plots the bias in $Q$ against the bias in $U$ over the 100 noise realizations for the band configuration with $\nu_{\rm min} = 30$\,GHz and $\nu_{\rm max} = 500$\,GHz. When the data is simulated using the MBB or HD models, the bias on the recovered CMB $Q$ and $U$ are roughly equal in each noise realization. For the fits to the Cloud model, however, the $Q$ and $U$ biases are related by a line of different slope and intercept, despite the input CMB signal having $Q = U$ at all frequencies. The Cloud model therefore directly introduces errors on the recovered CMB polarization angle.

\begin{figure}[t]
    \centering
	\includegraphics[width=0.48\textwidth]{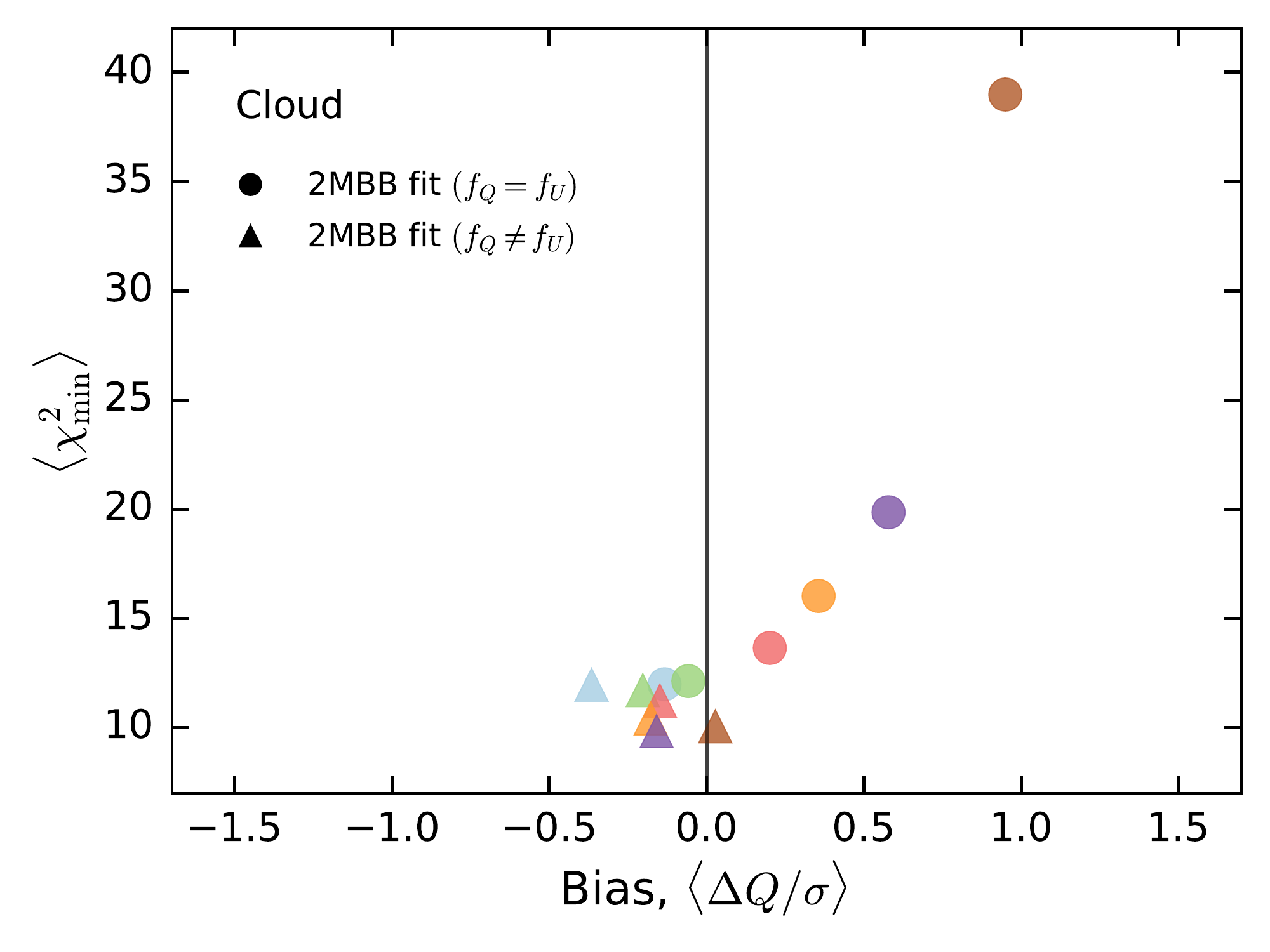}
	\caption{Mean bias in the CMB Q amplitude and median of the minimum $\chi^2$ for 2MBB fits to a Cloud input model, with $f_Q = f_U$ (circles) and $f_Q \neq f_U$ (triangles). The colors denote $\nu_{\rm max}$, as in Figure~\ref{fig:chi2_mbb}; only the $\nu_{\rm min} = 30$ GHz points are shown.}
	\label{fig:cloud_depol}
\end{figure}

The results of fitting the Cloud model with the more complex 2MBB model are also shown in Figs.~\ref{fig:chi2_mbb} and \ref{fig:bias_mbb}. As we noted in Section~\ref{sec:spfits}, this model assumes that the two dust components are aligned by the same magnetic field (i.e., $f_Q = f_U$) and thus have the same polarization angle. Due to this assumption, the 2MBB model is unable to account for the frequency-dependent polarization angle that arises in the Cloud model. Allowing $f_Q \neq f_U$ would solve this issue, at the expense of introducing one more free parameter -- see Figure~\ref{fig:cloud_depol} for a demonstration. As with the MBB fits, the ($f_Q = f_U$) 2MBB fits recover CMB amplitudes that are biased for all frequency configurations, with those having higher $\nu_{\rm max}$ also having the largest $\chi^2$. Overall, the bias is less than in the MBB fits due to the greater flexibility of the model.

While dust models with multiple components have been considered in the context of upcoming CMB experiments, the components are almost always assumed to be aligned by the same magnetic field and thus have the same polarization angle \citep[e.g.][]{ArmitageCaplan+etal_2012,Remazeilles+etal_2016,Thorne+etal_2017}. However, even for the simple case of a small temperature difference between clouds, we find that line of sight effects can induce non-negligible biases in the recovered CMB polarization at the noise levels considered. Additional complications, such as more severely misaligned clouds or different $\beta_d$ \apjedit{values} in each cloud, would exacerbate this effect.

\begin{figure*}[t]
    \centering
    \vspace{-2em}
	\includegraphics[width=0.78\textwidth]{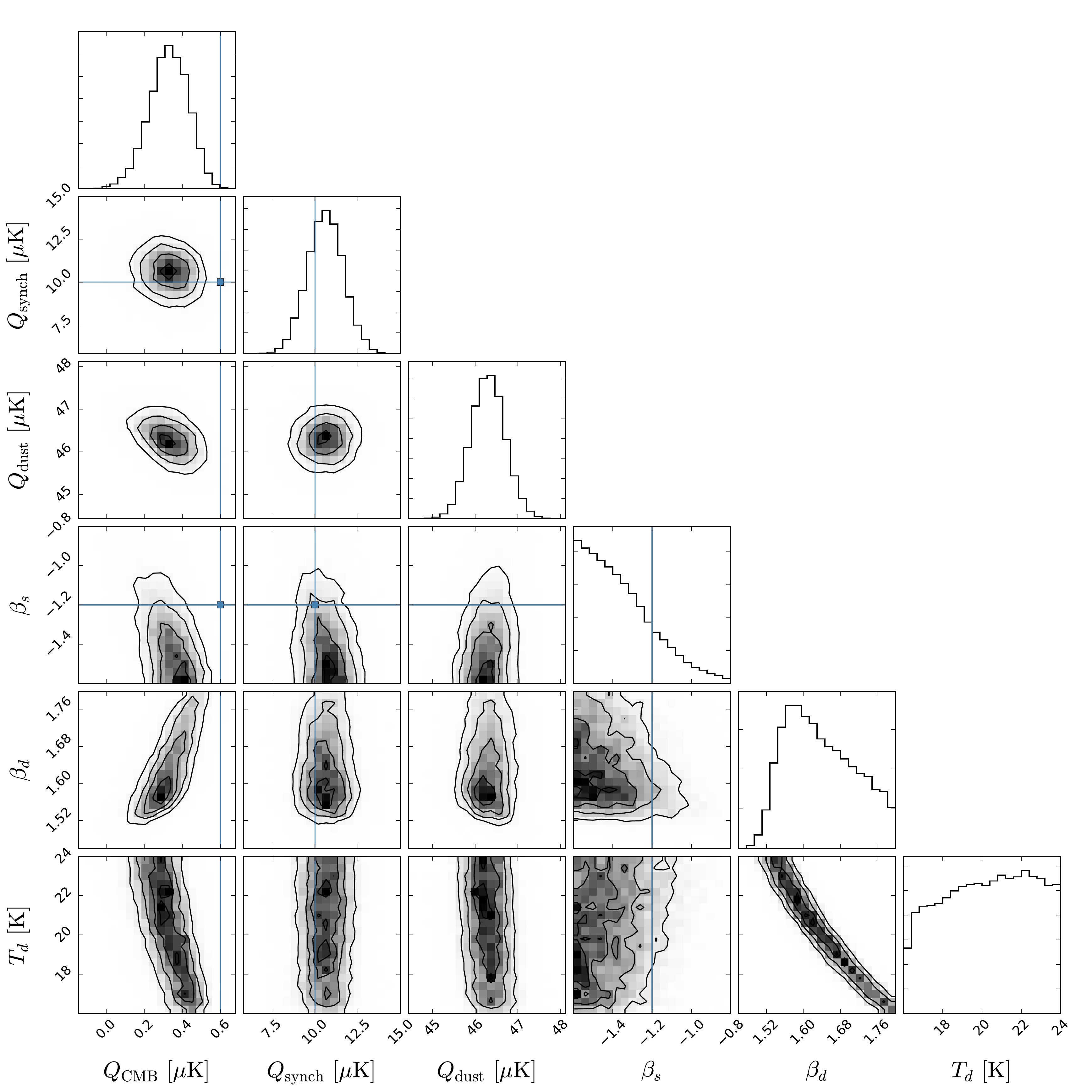}
	\caption{Posterior distributions of selected CMB, synchrotron, and dust parameters for an MBB fit to an Fe input model, for a single noise realization with $\nu_{\rm min}, \nu_{\rm max}= 30,\, 500$ GHz. The blue lines show the input values of parameters (where appropriate).}
	\label{fig:2mbbfe_chain}
\end{figure*}

\begin{figure}[t]
    \centering
	\includegraphics[width=0.48\textwidth]{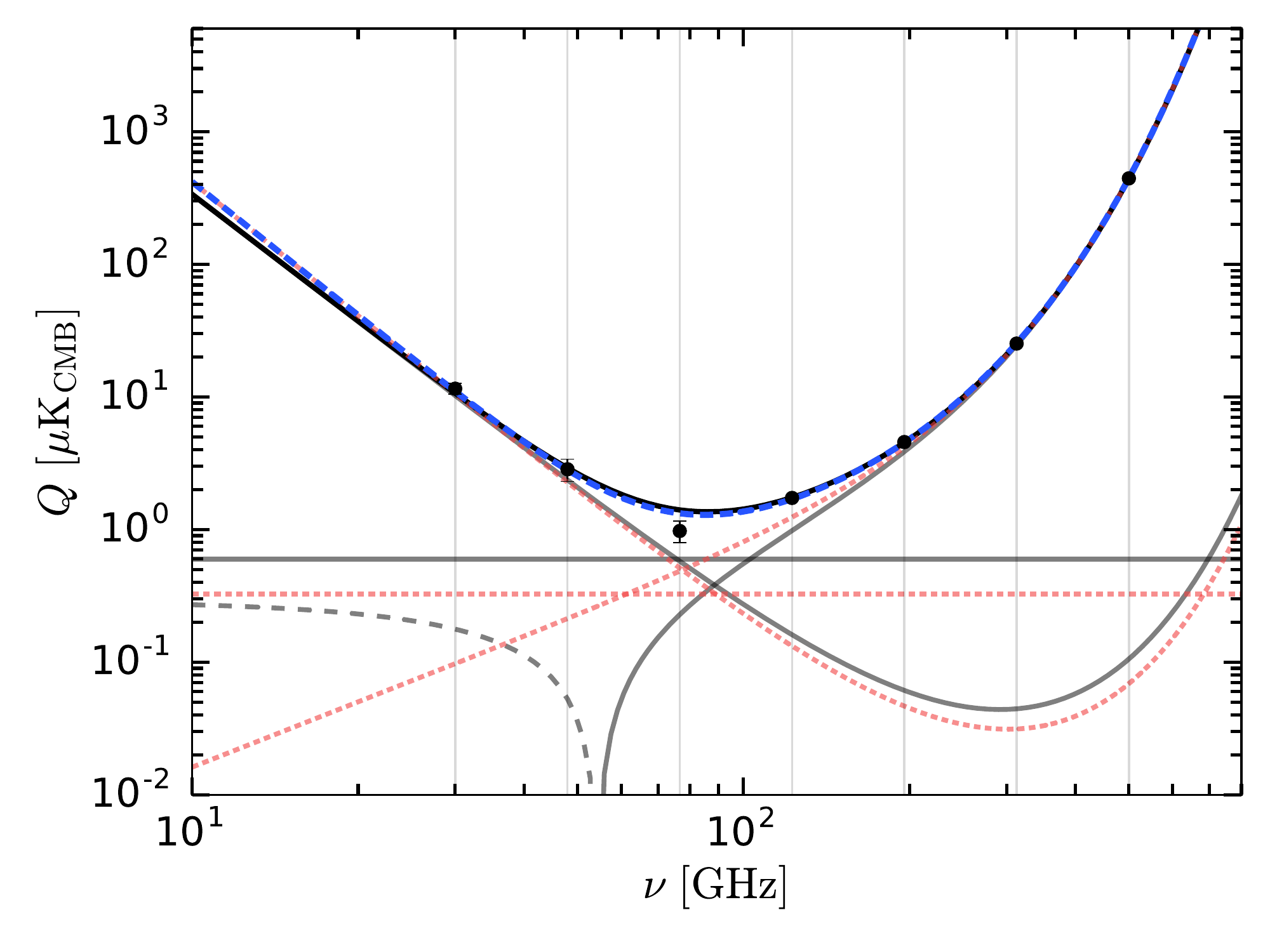}
	\caption{The SEDs of the CMB, synchrotron, and Fe dust components, for the input model (gray) and posterior mean values of the fitting parameters shown in Figure~\ref{fig:2mbbfe_chain} (red). The total SEDs are shown in black and blue respectively, while the simulated data are shown as points. The vertical gray lines show the location of each band.}
	\label{fig:2mbbfe_spectrum}
\end{figure}

\begin{table*}
\centering
 \setlength\tabcolsep{0pt}
\renewcommand*{\arraystretch}{1.6}
\begin{tabular}{L{1.6cm}|C{1.4cm}C{1.4cm}C{1.4cm}|C{1.4cm}C{1.4cm}C{1.4cm}|C{1.4cm}C{1.4cm}C{1.4cm}|}
   \multicolumn{1}{c}{} & \multicolumn{9}{c}{{\bf MBB fit}} \\
   & \multicolumn{3}{c|}{{\bf (20, 300) GHz}} & \multicolumn{3}{c|}{{\bf (20, 500) GHz}} & \multicolumn{3}{c|}{{\bf (20, 700) GHz}} \\
  {\bf } & $\langle \Delta Q / \sigma \rangle$ & $\langle \chi^2_{\rm min} \rangle$ & $\langle {\rm SNR}_Q \rangle$ & $\langle \Delta Q / \sigma \rangle$ & $\langle \chi^2_{\rm min} \rangle$ & $\langle {\rm SNR}_Q \rangle$ & $\langle \Delta Q / \sigma \rangle$ & $\langle \chi^2_{\rm min} \rangle$ & $\langle {\rm SNR}_Q \rangle$ \\
  \hline
MBB & \cellcolor{Red!5} $-0.15$ & \cellcolor{ForestGreen!9} $11.5$ & \cellcolor{ForestGreen!29} $4.0$ & \cellcolor{Cerulean!3} $+0.10$ & \cellcolor{ForestGreen!6} $11.4$ & \cellcolor{ForestGreen!40} $5.0$ & \cellcolor{Cerulean!3} $+0.10$ & \cellcolor{Red!2} $10.9$ & \cellcolor{ForestGreen!37} $4.7$ \\
SilCar & \cellcolor{Red!17} $-0.50$ & \cellcolor{ForestGreen!11} $11.7$ & \cellcolor{ForestGreen!32} $4.2$ & \cellcolor{Red!43} $-1.25$ & \cellcolor{ForestGreen!15} $11.9$ & \cellcolor{ForestGreen!33} $4.3$ & \cellcolor{Red!46} $-1.34$ & \cellcolor{ForestGreen!30} $12.7$ & \cellcolor{ForestGreen!29} $3.9$ \\
MF15 & \cellcolor{Red!17} $-0.51$ & \cellcolor{ForestGreen!26} $12.5$ & \cellcolor{ForestGreen!44} $5.5$ & \cellcolor{Red!57} $-1.64$ & \cellcolor{ForestGreen!40} $13.3$ & \cellcolor{ForestGreen!54} $6.5$ & \cellcolor{Red!56} $-1.60$ & \cellcolor{ForestGreen!38} $13.2$ & \cellcolor{ForestGreen!53} $6.4$ \\
Cloud & \cellcolor{Red!5} $-0.15$ & \cellcolor{ForestGreen!10} $11.6$ & \cellcolor{ForestGreen!31} $4.1$ & \cellcolor{Cerulean!2} $+0.06$ & \cellcolor{ForestGreen!41} $13.4$ & \cellcolor{ForestGreen!39} $5.0$ & \cellcolor{Cerulean!24} $+0.69$ & \cellcolor{ForestGreen!70} $20.4$ & \cellcolor{ForestGreen!41} $5.1$ \\
Fe & \cellcolor{Red!60} $-1.72$ & \cellcolor{ForestGreen!9} $11.6$ & \cellcolor{ForestGreen!13} $2.3$ & \cellcolor{Red!70} $-2.10$ & \cellcolor{ForestGreen!7} $11.5$ & \cellcolor{ForestGreen!21} $3.2$ & \cellcolor{Red!61} $-1.76$ & \cellcolor{Red!2} $10.8$ & \cellcolor{ForestGreen!18} $2.9$ \\
HD & \cellcolor{Cerulean!19} $+0.55$ & \cellcolor{ForestGreen!24} $12.4$ & \cellcolor{ForestGreen!55} $6.6$ & \cellcolor{Red!4} $-0.12$ & \cellcolor{ForestGreen!29} $12.7$ & \cellcolor{ForestGreen!67} $7.7$ & \cellcolor{Cerulean!31} $+0.90$ & \cellcolor{ForestGreen!28} $12.6$ & \cellcolor{ForestGreen!68} $7.8$ \\
HD + Fe & \cellcolor{Red!17} $-0.50$ & \cellcolor{ForestGreen!14} $11.9$ & \cellcolor{ForestGreen!37} $4.8$ & \cellcolor{Red!28} $-0.81$ & \cellcolor{ForestGreen!9} $11.5$ & \cellcolor{ForestGreen!40} $5.0$ & \cellcolor{Red!2} $-0.08$ & \cellcolor{ForestGreen!25} $12.5$ & \cellcolor{ForestGreen!51} $6.1$ \\

  \hline
 \multicolumn{1}{c}{} & \multicolumn{9}{c}{{\bf 2MBB fit}} \\
  {\bf } & $\langle \Delta Q / \sigma \rangle$ & $\langle \chi^2_{\rm min} \rangle$ & $\langle {\rm SNR}_Q \rangle$ & $\langle \Delta Q / \sigma \rangle$ & $\langle \chi^2_{\rm min} \rangle$ & $\langle {\rm SNR}_Q \rangle$ & $\langle \Delta Q / \sigma \rangle$ & $\langle \chi^2_{\rm min} \rangle$ & $\langle {\rm SNR}_Q \rangle$ \\
  \hline
MBB & \cellcolor{Red!8} $-0.23$ & \cellcolor{ForestGreen!13} $11.8$ & \cellcolor{ForestGreen!18} $2.8$ & \cellcolor{Red!5} $-0.17$ & \cellcolor{ForestGreen!5} $11.3$ & \cellcolor{ForestGreen!18} $2.8$ & \cellcolor{Cerulean!1} $+0.05$ & \cellcolor{Red!2} $10.8$ & \cellcolor{ForestGreen!22} $3.3$ \\
SilCar & \cellcolor{Red!15} $-0.43$ & \cellcolor{ForestGreen!12} $11.7$ & \cellcolor{ForestGreen!12} $2.2$ & \cellcolor{Red!9} $-0.28$ & \cellcolor{ForestGreen!8} $11.5$ & \cellcolor{ForestGreen!5} $1.6$ & \cellcolor{Cerulean!0} $+0.02$ & \cellcolor{Red!1} $10.9$ & \cellcolor{ForestGreen!18} $2.8$ \\
MF15 & \cellcolor{Red!21} $-0.62$ & \cellcolor{ForestGreen!17} $12.0$ & \cellcolor{ForestGreen!5} $1.6$ & \cellcolor{Red!13} $-0.39$ & \cellcolor{ForestGreen!4} $11.2$ & \cellcolor{ForestGreen!1} $1.1$ & \cellcolor{Red!14} $-0.43$ & \cellcolor{ForestGreen!0} $11.1$ & \cellcolor{ForestGreen!9} $2.0$ \\
Cloud & \cellcolor{Red!3} $-0.10$ & \cellcolor{ForestGreen!11} $11.7$ & \cellcolor{ForestGreen!16} $2.6$ & \cellcolor{Cerulean!4} $+0.12$ & \cellcolor{ForestGreen!39} $13.2$ & \cellcolor{ForestGreen!14} $2.4$ & \cellcolor{Cerulean!20} $+0.59$ & \cellcolor{ForestGreen!70} $20.3$ & \cellcolor{ForestGreen!25} $3.5$ \\
Fe & \cellcolor{Red!51} $-1.47$ & \cellcolor{ForestGreen!14} $11.9$ & \cellcolor{ForestGreen!6} $1.6$ & \cellcolor{Red!56} $-1.62$ & \cellcolor{ForestGreen!5} $11.3$ & \cellcolor{ForestGreen!5} $1.6$ & \cellcolor{Red!46} $-1.33$ & \cellcolor{Red!3} $10.8$ & \cellcolor{ForestGreen!9} $1.9$ \\
HD & \cellcolor{Red!5} $-0.15$ & \cellcolor{ForestGreen!13} $11.8$ & \cellcolor{ForestGreen!10} $2.1$ & \cellcolor{Cerulean!0} $+0.01$ & \cellcolor{ForestGreen!6} $11.4$ & \cellcolor{ForestGreen!5} $1.6$ & \cellcolor{Red!10} $-0.30$ & \cellcolor{Red!1} $10.9$ & \cellcolor{ForestGreen!9} $2.0$ \\
HD + Fe & \cellcolor{Red!16} $-0.48$ & \cellcolor{ForestGreen!14} $11.8$ & \cellcolor{ForestGreen!10} $2.0$ & \cellcolor{Red!12} $-0.34$ & \cellcolor{ForestGreen!3} $11.2$ & \cellcolor{ForestGreen!5} $1.6$ & \cellcolor{Red!23} $-0.68$ & \cellcolor{Red!2} $10.8$ & \cellcolor{ForestGreen!9} $2.0$ \\
  \hline
\end{tabular}
\caption{Summary statistics over 100 noise realizations for three selected band configurations, for all 7 input models and both fitting models. The summary statistics are the mean of the error-normalized bias for the CMB Q amplitude; the median of the minimum $\chi^2$; and the mean signal-to-noise ratio of the CMB Q amplitude.}
\label{tbl:stats}
\end{table*}

{\it Planck} has observed evidence of spatial variation in the spectrum of the polarized dust emission \citep{Planck_Int_L}. Depending on the magnitude of the effect, component separation methods relying on persistent spatial correlation of the dust across frequencies may be severely impacted, rendering parametric methods all the more important. Models such as the Cloud model analyzed here can help test these analysis methods against plausible line of sight effects, which we have demonstrated can be at levels of concern for CMB science.

\subsection{Models based on microscopic dust physics}
\label{sec:hd}

The models discussed so far have been instantiations of the generalized modified blackbody model described in Section~\ref{sec:gmbb}. While these models provide a means of investigating a number of physical effects with convenient analytic functions, they likely fail to account in detail for the intrinsic frequency-dependence of the dust emission arising from the long wavelength properties of amorphous materials. Indeed, laboratory studies have demonstrated that interstellar dust analogues can have diverse and complex opacity laws in the FIR that are not well described by simple power laws \citep{Agladze+etal_1996, Demyk+etal_2017a, Demyk+etal_2017b}. Further, the total microwave emission from dust arises from grains of different sizes, temperatures, and compositions. To evaluate the impact of these complications, we employ the physical dust model of \citet{Hensley+Draine_2017c} (HD), which was described in Section~\ref{sec:hd16}.

We begin by fitting the HD model simulations with the simple MBB model. When high frequencies are included ($\nu_{\rm max} \geq 700$\,GHz), the recovered CMB is biased by more than $1\sigma$, but the $\chi^2$ values are generally elevated, making it easy to identify the poor fit. The remaining configurations have relatively low bias ($< 0.5\sigma$) and acceptable $\chi^2$ values. The silicate opacity adopted by \citet{Hensley+Draine_2017c} is not a perfect power law, and thus fitting it as a power law over a wide frequency range induces modeling errors. Additionally, the dust SED is more sensitive to the underlying dust temperature distribution at higher frequencies. Thus, the model failures are more severe for those configurations with coverage extending to high frequencies.

Fits to the HD model with the 2MBB fitting function are presented in the right panels of Figs.~\ref{fig:chi2_mbb} and \ref{fig:bias_mbb}. In this case, all frequency configurations achieve an acceptable $\chi^2$ goodness of fit, but those with the highest frequency bands remain significantly biased. Thus, while the extra degrees of freedom provided by the 2MBB model allow a better fit to be obtained, the inferred parameters are not accurate, and give a misleading picture of the underlying set of components. While having bands at high frequencies is often useful for identifying model errors like this, as seen previously for the Cloud model (Section~\ref{sec:depol}), it also requires a reliable parametric model that is valid over a large range of frequencies. The 2MBB model fails in this respect, as it cannot reproduce the non-ideal behavior of the dust opacity in the HD model. Experiments employing high-frequency bands must therefore ensure that their analysis techniques are robust to this effect, as there is a significant risk that it will silently bias the recovered CMB amplitudes.

\subsection{Models with an iron grain component}
\label{sec:fe}

Metallic iron nanoparticles may be a significant component of interstellar dust with strong emissivity in the microwave \citep{Draine+Hensley_2013,Hensley+Draine_2017c}. This component is potentially problematic for CMB experiments due both to its relatively flat spectrum and its polarization signature. In particular, as metallic iron inclusions emit magnetic dipole radiation, their emission is polarized orthogonally to that of the grain in which they are embedded. Therefore, iron inclusions can induce a frequency-dependent change in the dust polarization fraction. We explore their effects in the contexts of two models: one in which the iron component is modeled as a simple graybody (i.e., $\beta_d = 0$; see Section~\ref{ssec:mbb_fe}) which we denote `Fe,' and a more physical model which also incorporates magnetic resonance effects \citep[][see Section~\ref{sec:hd16}]{Hensley+Draine_2017c}, which we denote `HD + Fe.'

In Figure~\ref{fig:chi2_mbb}, we present the results of fitting the Fe model with both the MBB (left) and 2MBB (right) models. In both cases and for all frequency configurations, the fits are biased by more than $1\sigma$. This is not surprising, as neither model can account for the two orthogonal polarization angles contributing to the total emission. More disconcerting, however, is the overall goodness of the fits in all cases, despite the strong bias. The flat spectrum of the iron grains is readily mimicked by the CMB, leading to substantial bias in the recovered CMB amplitudes. 

These effects are illustrated more clearly in Figures~\ref{fig:2mbbfe_chain} and \ref{fig:2mbbfe_spectrum}, which present the posterior distributions for select model parameters and the best-fit component SEDs, respectively, for an MBB fit to the Fe model with $\nu_{\rm min} = 30$\,GHz and $\nu_{\rm max} = 500$\,GHz. Figure~\ref{fig:2mbbfe_chain} shows that the chains have converged on a seemingly good best-fit model, and that the model parameters are well-determined, with no degeneracies apart from the usual $\beta_d - T_d$ degeneracy. Similarly, Figure~\ref{fig:2mbbfe_spectrum} shows that the total SED of the best-fit model (blue line) provides an excellent fit to the data, with only a slight deviation from one data point near 80\,GHz. Nevertheless, the posterior excludes the true CMB Q amplitude with high significance. This occurs because the decay and sign change of the dust SED induced by the iron component at low frequencies cannot be modeled by the MBB component used in the fits, but is readily compensated by a reduction in CMB amplitude and a shift in the synchrotron spectral index. In fact, with this choice of fitting model, there is no solution that can simultaneously recover the correct CMB amplitudes whilst also producing a good fit to the data -- the assumed MBB dust model is just too inflexible at low frequency. This degeneracy renders iron grains a potentially pernicious foreground for parametric component separation methods, as large biases can be induced without leaving any tell-tale signs, such as a poor goodness of fit.

In the bottom left panel of Figure~\ref{fig:chi2_mbb}, we present the results of fitting the HD + Fe model with the MBB model. In this case, we find that configurations with $\nu_{\rm max} \leq 600$\,GHz are somewhat biased ($\lesssim 1\sigma$) while having low $\chi^2$ values. In contrast, the two configurations with the highest $\nu_{\rm max}$ have essentially no bias but higher $\chi^2$. These latter configurations have fewer frequency bands covering the region where the iron emission and the CMB are both significant, and thus the degeneracy between the CMB and the magnetic emission is less severe. However, by having more bands dominated by dust emission that is more complicated than the simple MBB parameterization, the goodness of fit is poorer than for configurations with more bands at lower frequencies. In general, the biases are less than observed in the case of the `Fe' model as the iron component in this model contributes a smaller fraction of the total polarized signal (see Figure~\ref{fig:dust_models}).

Finally, in the bottom right panel of Figure~\ref{fig:chi2_mbb}, we present the results of fitting the HD + Fe model with the 2MBB fitting function. All frequency configurations have a similar and acceptable $\chi^2$. However, those with high $\nu_{\rm max}$ are significantly biased, whereas the bias is smaller for those experiments concentrated at lower frequencies. The extra degrees of freedom introduced by the 2MBB fit allow much, but not all, of the complexity of the HD + Fe model to be absorbed. Those experiments with high frequencies must correctly model the transition from the dust polarized emission being dominated by the silicate component to being dominated by the iron component. This cannot be described as the sum of two modified blackbodies having the same polarization angle. In contrast, the low frequency emission dominated by the iron inclusions can be well-described by a modified blackbody and thus subtracted more effectively. Any residual modeling errors at higher frequencies can be compensated by the second dust component while inducing only minimal bias at frequencies where the CMB dominates.

Iron grains pose a substantial challenge to component separation due to a spectrum that is degenerate with the CMB and their unique polarization signature. Sufficiently flexible models are required to fit this emission, but the additional complexity must be balanced against an increased ability to fit bad models and still achieve good fits. This problem can be mitigated by increasing the number of frequency bands.

\subsection{Causes of bias in the 2MBB fits}
\label{sec:bias_sources}

As discussed above, some of the models return biased results even when the significantly more flexible 2MBB model is used to fit the dust. This is not particularly surprising for the HD and HD + Fe models, which are more complicated than a simple superposition of modified blackbodies (meaning that model errors are expected), while our decision to set $f_Q = f_U$ means that the frequency decorrelation effects in the Cloud model cannot be fully captured. Nevertheless, there are two models that still produce significant biases even though the 2MBB model has the freedom to reproduce them exactly. These are the MF15 and Fe models, which result in $\sim 0.2-0.8\sigma$ and $\sim 1.0-1.5\sigma$ biases respectively, but always with acceptable $\chi^2$ values.

One potential cause of this behavior is the presence of degeneracies between dust model parameters. There is a well-known degeneracy between dust temperature and spectral index even in the MBB case, as illustrated in Figure~\ref{fig:2mbbfe_chain}. The more complicated 2MBB parameter space also supports several additional degeneracies, since changes in the parameters of one of the constituent MBB components can often be compensated by changes in the other. As such, entirely different choices of 2MBB parameters can sometimes give almost exactly the same SEDs. If one considers that there is also substantial freedom available in the fits from the CMB and synchrotron components, then it is clear that the parameter space in this fitting problem is likely to be quite complex, with the posteriors possibly exhibiting multiple local maxima. The risk of this happening is even greater for input models which have spectral features similar to the SEDs of other components (e.g. the flat low-frequency part of the Fe SED mimics the CMB). The existence of multiple maxima is problematic; while the correct, unbiased CMB amplitude may be found in one maximum, other maxima could give a strongly biased result, with no easy way to choose which one is correct.

We do not see obvious multi-modality in the bias histogram of Figure~\ref{fig:bias_mbb} however, and obtain similarly biased results when the chains are started at exactly the correct input parameter values. There is therefore a persistent preference for the incorrect, biased parameter values in some of the 2MBB fits. The recovered best-fit values can change significantly when different prior ranges are chosen for some of the parameters, however. For example, MCMC fits to the MF15 model with the same noise realization but using different prior ranges for $\beta_d$ resulted in very different best-fit values for the CMB Q amplitude. In our tests, we found biases ranging from $0.3\sigma$ for $\beta_d \in [1.4, 1.7]$, to $1.1\sigma$ for $\beta_d \in [1.4, 1.8]$, with results in between for different choices of minimum/maximum.

Based on this sensitivity to the priors on some parameters, we suspect that the real issue is that the flat priors we have chosen on the spectral parameters are actually informative. This effect has been pointed out previously in the context of the foreground component separation problem by \cite{EriksenCommander}. While flat priors are properly uninformative for linear parameters in the model (e.g., amplitudes or additive constants), they are informative for the nonlinear spectral parameters such as $\beta_d$, $\Delta\beta_d$, $T_{d,1}$, and $T_{d,2}$, i.e., some values of these parameters are preferred over others. This can result in biased posterior distributions. When the posterior has multiple maxima, the effect of an informative prior could be enough to single out one of the biased maxima as the global maximum, disfavoring the `true' maximum.

It should be possible to mitigate this effect by using a `Jeffreys prior,' which is proportional to the square root of the determinant of the Fisher matrix. This choice of prior is uninformative even for nonlinear parameters. However, we defer a more detailed study of how to avoid these biases to future work.

\subsection{Signal-to-noise optimization}
\label{sec:snrs}

So far we have focused on how different foreground modeling assumptions and band specifications affect the likelihood that the recovered CMB polarization will be biased, and whether the bias can be identified. These factors also have an important effect on the sensitivity of the measurements. Since component separation can never be performed perfectly due to the presence of instrumental noise, it will always result in some residual foreground signal being left in the estimated CMB map. This acts as an additional noise term, typically reducing the signal-to-noise ratio compared with an `ideal' scenario with no foregrounds. The component separation procedure can also affect the recovered signal in other ways, such as by over-subtracting the foregrounds (and therefore suppressing the true CMB signal), or by over-fitting, resulting in an artificial reduction in the noise level. As we have seen above, the effectiveness of the component separation procedure can depend strongly on the minimum and maximum frequencies of the bands that are available.

Some example SNR statistics are shown in Table~\ref{tbl:stats}, for three different band specifications and the two different fitting models. There is a clear dependence on the fitting model, with the MBB fits resulting in SNRs that can be factors of several larger than in the 2MBB fits. This can be understood by counting degrees of freedom -- since the 2MBB model has several more parameters than MBB, most of the information available in the data is being used to constrain those degrees of freedom instead of beating down the noise on the CMB amplitudes. In other words, marginalizing over more nuisance parameters (in this case, parameters of the dust model) reduces the precision with which the CMB signal can be recovered. This is problematic -- while simple MBB models are unable to model complex dust scenarios with sufficient accuracy, the more successful 2MBB models are considerably more complex, and result in a significant reduction in sensitivity to the CMB signal. A potential solution is to include several more bands in the instrument design, but we have not considered this possibility here.

Band configurations with higher $\nu_{\rm max}$ have a slight tendency to result in larger SNRs. This is primarily because the dust signal increases with frequency, and so one can gain a better handle on at least some of the dust parameters by increasing $\nu_{\rm max}$. This is not always the case however, as model errors caused by the increasing complexity of the dust spectra with frequency can result in the CMB signal being systematically underestimated, reducing the SNR. A similar effect can also be seen at low frequency for the Fe model, which has a flattened spectrum below $\sim 100$ GHz that introduces a degeneracy with the primary CMB. Degeneracies can also lead to systematic {\it over}estimates of the CMB amplitude, which probably explains why fitting a simple MBB model to the HD model results in a higher SNR than when fitting the MBB model to itself.

As noted previously, increasing $\nu_{\rm max}$ can reduce the risk of bias in some scenarios, while increasing it in others. As such, there is not really an `optimal' band configuration that is robust against any plausible dust complexity. This conclusion carries through when one also considers the SNR -- the optimal band configuration depends on what the true dust model is, which we don't know. For a given model the SNR is only mildly sensitive to $\nu_{\rm max}$ however, suggesting that future experiments should focus on finding configurations that reduce bias first, and then optimize for SNR as a secondary concern.

\subsection{Interpretation of bias statistics}
\label{sec:bias_interp}

In order to study a wide range of possible instrumental configurations and dust scenarios (each with a large set of noise realizations), we have restricted ourselves to a single-pixel analysis. The lack of full-sky simulations means that we are unable to translate our findings into implications for measurements of $r$, the tensor-to-scalar ratio. This is left for future work, which will be able to consider a more restricted set of scenarios informed by the findings in this paper.

Nevertheless, we can make some simplistic statements about the possible effects of the biases that we have identified on a cosmological parameter analysis. Comparing the most idealistic case -- MBB fits to a true MBB dust model -- with the other scenarios, we see that the expected SNR on the polarized CMB does not change by much except for in the case of the Fe model (see Section~\ref{sec:snrs}). As such, the basic sensitivity to $r$ is not expected to change by much compared to the simulated MBB component separation analyses that have been performed previously \citep[e.g.][]{Alonso+etal_2017}. The problem in this case is the bias, however. For the 2MBB fits, the SNR per pixel is generally at least a factor of 2 worse, which translates to a factor of 4 on the CMB power spectrum.

It is difficult to estimate the effect of the bias without making maps of the residuals, as the contamination of the B-mode signal depends on the pattern of the residuals on the sky, i.e. how much of the residual is in B-modes as opposed to E-modes. Still, a bias of $1 \sigma$ per pixel means that the power in residual foregrounds is comparable to $\sim 1 / {\rm SNR}^2 \approx 1/5^2$ of the power in the CMB polarization on pixel scales. We used pixels of size $\sim 1^\circ$, corresponding to $\ell \sim 200$, where the CMB EE power is $D_{200}^{\rm EE} \approx 1 \mu{\rm K}^2$. If one third of the power of the foreground residuals is in B-modes \citep{Planck_Int_XXX}, and they have an angular power spectrum similar to the dust itself ($D_\ell \sim \ell^{\,-0.5}$), this would translate to a B-mode contamination of order $\sim 0.1 \mu{\rm K}^2$ in angular power around the reionization feature at $\ell \sim 5$. This is about 3 orders of magnitude larger than the CMB BB power if $r = 10^{-2}$. While we caution against taking this back-of-the-envelope estimate too seriously, typical biases of $1\sigma$ per pixel would clearly be a serious matter for forthcoming CMB polarization experiments.

\begin{figure}[t]
    \centering
    \vspace{0.5em}
	\includegraphics[width=0.48\textwidth]{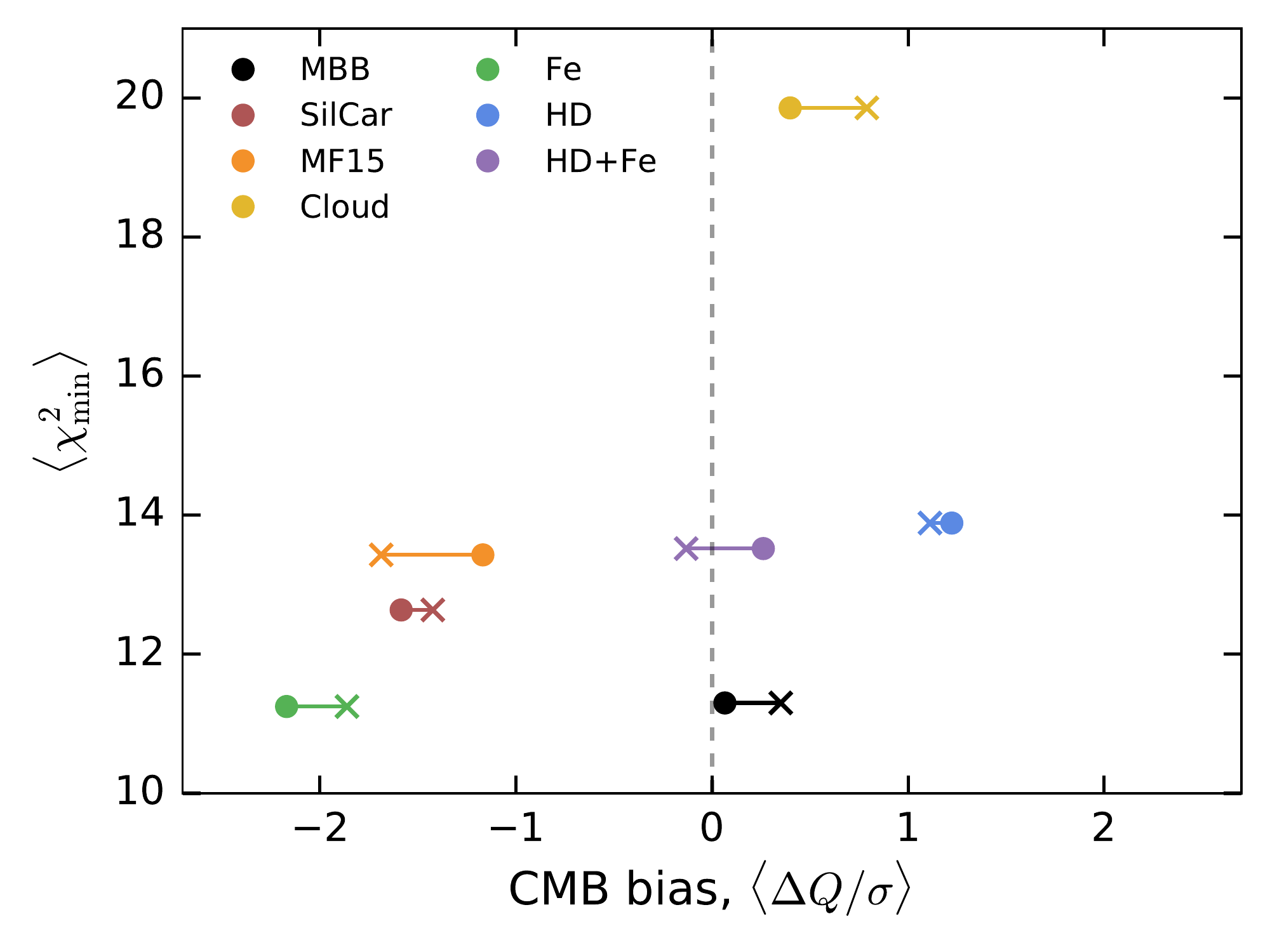}
	\caption{The mean bias in the CMB Q amplitude over 100 noise realizations, shown for two different bias definitions, plotted against the median of the minimum $\chi^2$. The two bias definitions use the mean (circles) and maximum a posteriori (crosses) values of the marginal posterior distribution for Q from each noise realization. The results are shown for all seven input dust models, assuming polarization-only MBB fits with $\nu_{\rm min}, \nu_{\rm max} = 30, 700$ GHz.}
	\label{fig:bias_summary_stats}
\end{figure}


Finally, we note that the precision of our determinations of the mean bias is limited by a couple of effects, of roughly similar magnitude. The first is the limited number of noise realizations that were used to estimate the mean. For a Gaussian random variable, the standard error on the mean scales like $\sigma / \sqrt{N}$, where $N$ is the number of realizations. We therefore expect an error on the mean bias of $\sim 0.1 \sigma$ for the standard choice of 100 noise realizations. (The number of realizations could be increased, but this would be computationally expensive -- the full set of results presented here required several days of wall clock time on 7x64 cores of a shared cluster.) Secondly, we have used the maximum a posteriori (MAP) estimate of the CMB Q and U amplitudes to define the bias, as discussed in Section~\ref{sec:spfits}. While this seems like the most sensible quantity to use, we could have also chosen other summary statistics, such as the mean. Figure~\ref{fig:bias_summary_stats} shows how the mean bias changes depending on whether the mean or MAP estimates are used for the bias definition. The difference between the two is around $0.1 - 0.5 \sigma$.

\section{Discussion and Conclusions}
\label{sec:discussion}

In this paper, we studied the ability of future CMB polarization experiments to remove foreground contamination that includes realistic, physically-motivated dust components. Until recently, most studies have assumed that galactic dust emission in the microwave band can be well-described by a simple modified blackbody form with only two spectral parameters. There are many reasons to expect that this may not be the case, however. In Section~\ref{sec:foreground} we enumerated some of the possibilities based on recent ab initio attempts to model galactic dust. These include:
\begin{itemize}
 \item Separate silicate and carbonaceous dust populations that have significantly different SEDs in temperature and polarization, so that a single MBB model (with spectral parameters that are the same in the Stokes I, Q, and U channels) cannot provide a good description;
 \item Models with line of sight effects (`cloud'), where the superposition of multiple dust clouds along the line of sight, each with different SEDs and polarization angles, adds complexity to the polarization spectrum, while leaving the total intensity unchanged;
 \item The presence of ferromagnetic iron grains, which give rise to magnetic dipole emission that produces flat, CMB-like dust SEDs at low frequencies;
 \item More complex `physical' dust models, based on modeling the detailed properties (e.g., size and temperature distributions) and relative abundances of different types of dust grains.
\end{itemize}
We found that these more realistic dust models can cause significant complications for attempts to remove dust foreground emission and therefore recover the true polarized CMB signal. In particular, biases in the CMB Q and U amplitudes can easily be larger than one standard deviation per $1^\circ$ pixel when simple MBB models are fitted to polarization-only data. These biases can often be identified by the enlarged residuals (poor $\chi^2$ fits) that they cause, with some important exceptions. For instance, employing MBB fits, we found that Silicate + Carbonaceous, MF15 \citep{Meisner+Finkbeiner_2015}, and iron-bearing (`Fe') dust models can result in $\sim 1-2 \sigma$ biases per pixel while still returning apparently good fits. This is dangerous, even if a good fraction of the residual dust contamination can be removed (e.g., by fitting an angular power spectrum template for the residual during cosmological parameter estimation).

This situation improves somewhat when a more sophisticated two-component MBB model is used. This can model a much wider range of dust physics, including the cloud, iron-bearing, and Silicate + Carbonaceous scenarios mentioned above, at the cost of introducing 3 or 4 additional parameters for a polarization-only analysis, depending on whether the polarization angle is allowed to change with frequency. There are several instances in which the recovered CMB amplitudes remain biased however. For the Cloud and HD models, the bias can be avoided by restricting the analysis to a relatively low maximum frequency, $\nu_{\rm max} \lesssim 400$ GHz. This is not a general solution, however: some models, like MF15 and especially Fe, remain significantly biased while still producing a good fit to the data (i.e., low $\chi^2$), regardless of the choice of minimum and maximum frequency. This is a serious cause for concern, as the bias would potentially be difficult to detect. Additionally, the 2MBB parameter space is more complicated than for MBB, and so extra care is needed to handle degeneracies, multi-modality, and biasing due to informative priors.

Contrary to expectations, using temperature data to augment the polarized foreground fits was not helpful even when making the simplifying assumption that the components have the same spectral parameters in both temperature and polarization. We explored this effect in Section~\ref{sec:temp_pol}, where we found that the additional complexity of the low-frequency temperature foregrounds (i.e., including AME and free-free emission) gives rise to degeneracies that have a knock-on effect on the CMB, causing small biases even when the correct (input) dust model is used for the fits. These biases can be mitigated somewhat by including frequency channels below $\sim 30$\,GHz, which help to break degeneracies between the low-frequency temperature components. We note, however, that our assumed AME and synchrotron models are highly idealized, neglecting, for instance, the line of sight effects explored in the dust emission which are equally applicable for the low frequency foregrounds. Multiple sub-30\,GHz bands will likely be required to account for the additional spectral degrees of freedom needed to provide a more realistic description of these components.

We did not consider possible spatial variations of foreground spectral parameters in this paper. Instead, we fixed the input parameter values to what should be reasonably typical values at intermediate to high latitudes. Future work will consider the effects of spatial variations, in particular frequency decorrelation arising from line of sight effects (i.e., due to clouds with a range of different SED parameters being averaged into one beam pointing). Similarly, by performing only a single pixel analysis, we have excluded the possibility of using morphological information to break degeneracies and identify biases. Dust models that leave significant residuals while still having apparently reasonable goodness-of-fit statistics are certainly dangerous in a single pixel context, but could give rise to residuals with tell-tale morphological signatures that are easier to identify. A full-sky analysis will be needed to understand whether this is the case or not.

Our conclusions are as follows:
\begin{enumerate}
 \item A single-temperature modified blackbody model of dust emission, even with spatially-varying spectral parameters, does not span the range of possible dust physics to which future CMB experiments will be sensitive. We have provided a library of physically-realistic dust models in Section~\ref{sec:thdust} that covers a variety of potential complexities. Some subset of these can be used to probe important dust physics effects, such as frequency decorrelation and the presence of ferromagnetic iron grains.
 \item High-quality multi-band data are needed to properly constrain the complex foregrounds in Stokes I at low frequencies. Even in the case of idealized low frequency foregrounds considered in this work, frequency bands below 30\,GHz are required to successfully disentangle the contributions of synchrotron, AME, free-free, and the CMB in total intensity. Curvature in the synchrotron SED, variations in the AME SED, and line of sight effects are all expected in real data and would exacerbate these degeneracies.
 \item Polarization-only analyses using a generalized two-component MBB (2MBB) model are the most robust fitting options that we considered, and should be regarded as a definite improvement to MBB in future analyses. The 2MBB models can still result in model errors and biases however. Restricting the analysis to relatively low maximum frequencies, $\nu_{\rm max} \sim 200-500$\,GHz can reduce the risk of bias in many scenarios, as some dust complexities become more severe at high frequencies. On the other hand, we found that high frequency bands ($\nu \gtrsim 500$\,GHz) can be critical for identifying poor model fits in other cases that result in biases on the recovered CMB amplitudes. A detailed study of these trade-offs should be undertaken in the context of specific proposed experiment designs. In particular, future work should examine the information gained by increasing the number of frequency channels.
 \item The 2MBB model can be further improved by allowing the two dust components to be polarized with different polarization ($f_Q \neq f_U$ in our notation). Although it introduces an extra parameter into the fit, this more general model is better able to capture the line of sight effects that can lead to frequency decorrelation.
\end{enumerate}

\acknowledgments
{The Python code and data files for the summary statistics calculated as part of this analysis are available to download from \url{http://philbull.com/singlepixel/}.

We are grateful to Geoff Bryden, Curt Cutler, Clive Dickinson, Tim Eifler, Hans Kristian Eriksen, Charles Lawrence, Graca Rocha, and especially Jeff Jewell, for valuable discussions and comments. \apjedit{We thank the anonymous referee for feedback which improved the manuscript.}
This research was carried out at the Jet Propulsion Laboratory, California Institute of Technology, under a contract with the National Aeronautics and Space Administration. PB's research was supported by an appointment to the NASA Postdoctoral Program at the Jet Propulsion Laboratory, California Institute of Technology, administered by Universities Space Research Association under contract with NASA.}

\bibliography{cmbpol}

\begin{thebibliography}{}
\expandafter\ifx\csname natexlab\endcsname\relax\def\natexlab#1{#1}\fi

\bibitem[{{Abazajian} {et~al.}(2016){Abazajian}, {Adshead}, {Ahmed}, {Allen},
  {Alonso}, {Arnold}, {Baccigalupi}, {Bartlett}, {Battaglia}, {Benson},
  {Bischoff}, {Borrill}, {Buza}, {Calabrese}, {Caldwell}, {Carlstrom}, {Chang},
  {Crawford}, {Cyr-Racine}, {De Bernardis}, {de Haan}, {di Serego Alighieri},
  {Dunkley}, {Dvorkin}, {Errard}, {Fabbian}, {Feeney}, {Ferraro}, {Filippini},
  {Flauger}, {Fuller}, {Gluscevic}, {Green}, {Grin}, {Grohs}, {Henning},
  {Hill}, {Hlozek}, {Holder}, {Holzapfel}, {Hu}, {Huffenberger}, {Keskitalo},
  {Knox}, {Kosowsky}, {Kovac}, {Kovetz}, {Kuo}, {Kusaka}, {Le Jeune}, {Lee},
  {Lilley}, {Loverde}, {Madhavacheril}, {Mantz}, {Marsh}, {McMahon},
  {Meerburg}, {Meyers}, {Miller}, {Munoz}, {Nguyen}, {Niemack}, {Peloso},
  {Peloton}, {Pogosian}, {Pryke}, {Raveri}, {Reichardt}, {Rocha}, {Rotti},
  {Schaan}, {Schmittfull}, {Scott}, {Sehgal}, {Shandera}, {Sherwin}, {Smith},
  {Sorbo}, {Starkman}, {Story}, {van Engelen}, {Vieira}, {Watson}, {Whitehorn},
  \& {Kimmy Wu}}]{CMBS4}
{Abazajian}, K.~N., {Adshead}, P., {Ahmed}, Z., {et~al.} 2016, ArXiv e-prints,
  arXiv:1610.02743

\bibitem[{{Agladze} {et~al.}(1996){Agladze}, {Sievers}, {Jones}, {Burlitch}, \&
  {Beckwith}}]{Agladze+etal_1996}
{Agladze}, N.~I., {Sievers}, A.~J., {Jones}, S.~A., {Burlitch}, J.~M., \&
  {Beckwith}, S.~V.~W. 1996, \apj, 462, 1026

\bibitem[{{Ali-Ha{\"i}moud} {et~al.}(2009){Ali-Ha{\"i}moud}, {Hirata}, \&
  {Dickinson}}]{Ali-Haimoud+Hirata+Dickinson_2009}
{Ali-Ha{\"i}moud}, Y., {Hirata}, C.~M., \& {Dickinson}, C. 2009, \mnras, 395,
  1055

\bibitem[{{Alonso} {et~al.}(2017){Alonso}, {Dunkley}, {Thorne}, \&
  {N{\ae}ss}}]{Alonso+etal_2017}
{Alonso}, D., {Dunkley}, J., {Thorne}, B., \& {N{\ae}ss}, S. 2017, \prd, 95,
  043504

\bibitem[{{Altobelli} {et~al.}(2016){Altobelli}, {Postberg}, {Fiege},
  {Trieloff}, {Kimura}, {Sterken}, {Hsu}, {Hillier}, {Khawaja},
  {Moragas-Klostermeyer}, {Blum}, {Burton}, {Srama}, {Kempf}, \&
  {Gruen}}]{Altobelli+etal_2016}
{Altobelli}, N., {Postberg}, F., {Fiege}, K., {et~al.} 2016, Science, 352, 312

\bibitem[{{Armitage-Caplan} {et~al.}(2012){Armitage-Caplan}, {Dunkley},
  {Eriksen}, \& {Dickinson}}]{ArmitageCaplan+etal_2012}
{Armitage-Caplan}, C., {Dunkley}, J., {Eriksen}, H.~K., \& {Dickinson}, C.
  2012, \mnras, 424, 1914

\bibitem[{{BICEP2 Collaboration} {et~al.}(2014){BICEP2 Collaboration}, {Ade},
  {Aikin}, {Barkats}, {Benton}, {Bischoff}, {Bock}, {Brevik}, {Buder},
  {Bullock}, {Dowell}, {Duband}, {Filippini}, {Fliescher}, {Golwala},
  {Halpern}, {Hasselfield}, {Hildebrandt}, {Hilton}, {Hristov}, {Irwin},
  {Karkare}, {Kaufman}, {Keating}, {Kernasovskiy}, {Kovac}, {Kuo}, {Leitch},
  {Lueker}, {Mason}, {Netterfield}, {Nguyen}, {O'Brient}, {Ogburn}, {Orlando},
  {Pryke}, {Reintsema}, {Richter}, {Schwarz}, {Sheehy}, {Staniszewski},
  {Sudiwala}, {Teply}, {Tolan}, {Turner}, {Vieregg}, {Wong}, \&
  {Yoon}}]{2014PhRvL.112x1101B}
{BICEP2 Collaboration}, {Ade}, P.~A.~R., {Aikin}, R.~W., {et~al.} 2014,
  Physical Review Letters, 112, 241101

\bibitem[{{BICEP2/Keck Collaboration} {et~al.}(2015){BICEP2/Keck
  Collaboration}, {Planck Collaboration}, {Ade}, {Aghanim}, {Ahmed}, {Aikin},
  {Alexander}, {Arnaud}, {Aumont}, {Baccigalupi}, \&
  et~al.}]{2015PhRvL.114j1301B}
{BICEP2/Keck Collaboration}, {Planck Collaboration}, {Ade}, P.~A.~R., {et~al.}
  2015, Physical Review Letters, 114, 101301

\bibitem[{{Chiar} {et~al.}(2006){Chiar}, {Adamson}, {Whittet}, {Chrysostomou},
  {Hough}, {Kerr}, {Mason}, {Roche}, \& {Wright}}]{Chiar+etal_2006}
{Chiar}, J.~E., {Adamson}, A.~J., {Whittet}, D.~C.~B., {et~al.} 2006, \apj,
  651, 268

\bibitem[{{Chluba} {et~al.}(2017){Chluba}, {Hill}, \&
  {Abitbol}}]{Chluba+Hill+Abitbol_2017}
{Chluba}, J., {Hill}, J.~C., \& {Abitbol}, M.~H. 2017, \mnras, 472, 1195

\bibitem[{{Demyk} {et~al.}(2017{\natexlab{a}}){Demyk}, {Meny}, {Leroux},
  {Depecker}, {Brubach}, {Roy}, {Nayral}, \& {Ojo}}]{Demyk+etal_2017b}
{Demyk}, K., {Meny}, C., {Leroux}, H., {et~al.} 2017{\natexlab{a}}, ArXiv
  e-prints, arXiv:1706.09801

\bibitem[{{Demyk} {et~al.}(2017{\natexlab{b}}){Demyk}, {Meny}, {Lu},
  {Papatheodorou}, {Toplis}, {Leroux}, {Depecker}, {Brubach}, {Roy}, {Nayral},
  {Ojo}, {Delpech}, {Paradis}, \& {Gromov}}]{Demyk+etal_2017a}
{Demyk}, K., {Meny}, C., {Lu}, X.-H., {et~al.} 2017{\natexlab{b}}, \aap, 600,
  A123

\bibitem[{{Dickinson} {et~al.}(2011){Dickinson}, {Peel}, \&
  {Vidal}}]{Dickinson+Peel+Vidal_2011}
{Dickinson}, C., {Peel}, M., \& {Vidal}, M. 2011, \mnras, 418, L35

\bibitem[{{Draine} \& {Hensley}(2012)}]{Draine+Hensley_2012}
{Draine}, B.~T., \& {Hensley}, B. 2012, \apj, 757, 103

\bibitem[{{Draine} \& {Hensley}(2013)}]{Draine+Hensley_2013}
---. 2013, \apj, 765, 159

\bibitem[{{Draine} \& {Hensley}(2016)}]{Draine+Hensley_2016}
{Draine}, B.~T., \& {Hensley}, B.~S. 2016, \apj, 831, 59

\bibitem[{{Draine} \& {Lazarian}(1998)}]{Draine+Lazarian_1998b}
{Draine}, B.~T., \& {Lazarian}, A. 1998, \apj, 508, 157

\bibitem[{{Draine} \& {Li}(2007)}]{Draine+Li_2007}
{Draine}, B.~T., \& {Li}, A. 2007, \apj, 657, 810

\bibitem[{{Eriksen} {et~al.}(2008){Eriksen}, {Jewell}, {Dickinson}, {Banday},
  {G{\'o}rski}, \& {Lawrence}}]{EriksenCommander}
{Eriksen}, H.~K., {Jewell}, J.~B., {Dickinson}, C., {et~al.} 2008, \apj, 676,
  10

\bibitem[{{Finkbeiner} {et~al.}(1999){Finkbeiner}, {Davis}, \&
  {Schlegel}}]{Finkbeiner+Davis+Schlegel_1999}
{Finkbeiner}, D.~P., {Davis}, M., \& {Schlegel}, D.~J. 1999, \apj, 524, 867

\bibitem[{{Foreman-Mackey} {et~al.}(2013){Foreman-Mackey}, {Hogg}, {Lang}, \&
  {Goodman}}]{Foreman-Mackey+etal_2013}
{Foreman-Mackey}, D., {Hogg}, D.~W., {Lang}, D., \& {Goodman}, J. 2013, \pasp,
  125, 306

\bibitem[{{G{\'e}nova-Santos} {et~al.}(2017){G{\'e}nova-Santos},
  {Rubi{\~n}o-Mart{\'{\i}}n}, {Pel{\'a}ez-Santos}, {Poidevin}, {Rebolo},
  {Vignaga}, {Artal}, {Harper}, {Hoyland}, {Lasenby},
  {Mart{\'{\i}}nez-Gonz{\'a}lez}, {Piccirillo}, {Tramonte}, \&
  {Watson}}]{GenovaSantos+etal_2017}
{G{\'e}nova-Santos}, R., {Rubi{\~n}o-Mart{\'{\i}}n}, J.~A.,
  {Pel{\'a}ez-Santos}, A., {et~al.} 2017, \mnras, 464, 4107

\bibitem[{{G{\'o}rski} {et~al.}(2005){G{\'o}rski}, {Hivon}, {Banday},
  {Wandelt}, {Hansen}, {Reinecke}, \& {Bartelmann}}]{Gorski+etal_2005}
{G{\'o}rski}, K.~M., {Hivon}, E., {Banday}, A.~J., {et~al.} 2005, \apj, 622,
  759

\bibitem[{{Hensley} \& {Draine}(2017{\natexlab{a}})}]{Hensley+Draine_2017c}
{Hensley}, B., \& {Draine}, B.~T. 2017{\natexlab{a}}, In preparation

\bibitem[{{Hensley} \& {Draine}(2017{\natexlab{b}})}]{Hensley+Draine_2017b}
{Hensley}, B.~S., \& {Draine}, B.~T. 2017{\natexlab{b}}, \apj, 836, 179

\bibitem[{{Hoang} \& {Lazarian}(2016)}]{Hoang+Lazarian_2016}
{Hoang}, T., \& {Lazarian}, A. 2016, \apj, 831, 159

\bibitem[{{Jones} {et~al.}(2017){Jones}, {K{\"o}hler}, {Ysard}, {Bocchio}, \&
  {Verstraete}}]{Jones+etal_2017}
{Jones}, A.~P., {K{\"o}hler}, M., {Ysard}, N., {Bocchio}, M., \& {Verstraete},
  L. 2017, \aap, 602, A46

\bibitem[{{Keating} {et~al.}(1998){Keating}, {Timbie}, {Polnarev}, \&
  {Steinberger}}]{Keating+etal_1998}
{Keating}, B., {Timbie}, P., {Polnarev}, A., \& {Steinberger}, J. 1998, \apj,
  495, 580

\bibitem[{{Kogut} \& {Fixsen}(2016)}]{Kogut+Fixsen_2016}
{Kogut}, A., \& {Fixsen}, D.~J. 2016, \apj, 826, 101

\bibitem[{{Kogut} {et~al.}(2007){Kogut}, {Dunkley}, {Bennett}, {Dor{\'e}},
  {Gold}, {Halpern}, {Hinshaw}, {Jarosik}, {Komatsu}, {Nolta}, {Odegard},
  {Page}, {Spergel}, {Tucker}, {Weiland}, {Wollack}, \&
  {Wright}}]{Kogut+etal_2007}
{Kogut}, A., {Dunkley}, J., {Bennett}, C.~L., {et~al.} 2007, \apj, 665, 355

\bibitem[{{Li} \& {Draine}(2001)}]{Li+Draine_2001}
{Li}, A., \& {Draine}, B.~T. 2001, \apj, 554, 778

\bibitem[{{Macellari} {et~al.}(2011){Macellari}, {Pierpaoli}, {Dickinson}, \&
  {Vaillancourt}}]{Macellari+etal_2011}
{Macellari}, N., {Pierpaoli}, E., {Dickinson}, C., \& {Vaillancourt}, J.~E.
  2011, \mnras, 418, 888

\bibitem[{{Mathis} {et~al.}(1983){Mathis}, {Mezger}, \&
  {Panagia}}]{Mathis+Mezger+Panagia_1983}
{Mathis}, J.~S., {Mezger}, P.~G., \& {Panagia}, N. 1983, \aap, 128, 212

\bibitem[{{Matsumura} {et~al.}(2014){Matsumura}, {Akiba}, {Borrill}, {Chinone},
  {Dobbs}, {Fuke}, {Ghribi}, {Hasegawa}, {Hattori}, {Hattori}, {Hazumi},
  {Holzapfel}, {Inoue}, {Ishidoshiro}, {Ishino}, {Ishitsuka}, {Karatsu},
  {Katayama}, {Kawano}, {Kibayashi}, {Kibe}, {Kimura}, {Kimura}, {Koga},
  {Kozu}, {Komatsu}, {Lee}, {Matsuhara}, {Mima}, {Mitsuda}, {Mizukami},
  {Morii}, {Morishima}, {Murayama}, {Nagai}, {Nagata}, {Nakamura}, {Naruse},
  {Natsume}, {Nishibori}, {Nishino}, {Noda}, {Noguchi}, {Ogawa}, {Oguri},
  {Ohta}, {Otani}, {Richards}, {Sakai}, {Sato}, {Sato}, {Sekimoto}, {Shimizu},
  {Shinozaki}, {Sugita}, {Suzuki}, {Suzuki}, {Tajima}, {Takada}, {Takakura},
  {Takei}, {Tomaru}, {Uzawa}, {Wada}, {Watanabe}, {Yoshida}, {Yamasaki},
  {Yoshida}, \& {Yotsumoto}}]{LiteBIRD}
{Matsumura}, T., {Akiba}, Y., {Borrill}, J., {et~al.} 2014, Journal of Low
  Temperature Physics, 176, 733

\bibitem[{{Meisner} \& {Finkbeiner}(2015)}]{Meisner+Finkbeiner_2015}
{Meisner}, A.~M., \& {Finkbeiner}, D.~P. 2015, \apj, 798, 88

\bibitem[{{Miville-Desch{\^e}nes} {et~al.}(2008){Miville-Desch{\^e}nes},
  {Ysard}, {Lavabre}, {Ponthieu}, {Mac{\'{\i}}as-P{\'e}rez}, {Aumont}, \&
  {Bernard}}]{MivilleDeschenes+etal_2008}
{Miville-Desch{\^e}nes}, M.-A., {Ysard}, N., {Lavabre}, A., {et~al.} 2008,
  \aap, 490, 1093

\bibitem[{{Planck Collaboration} {et~al.}(2014{\natexlab{a}}){Planck
  Collaboration}, {Abergel}, {Ade}, {Aghanim}, {Alves}, {Aniano},
  {Armitage-Caplan}, {Arnaud}, {Ashdown}, {Atrio-Barandela}, \&
  et~al.}]{Planck_2013_XI}
{Planck Collaboration}, {Abergel}, A., {Ade}, P.~A.~R., {et~al.}
  2014{\natexlab{a}}, \aap, 571, A11

\bibitem[{{Planck Collaboration} {et~al.}(2014{\natexlab{b}}){Planck
  Collaboration}, {Ade}, {Aghanim}, {Armitage-Caplan}, {Arnaud}, {Ashdown},
  {Atrio-Barandela}, {Aumont}, {Baccigalupi}, {Banday}, \&
  et~al.}]{Planck_2013_XII}
{Planck Collaboration}, {Ade}, P.~A.~R., {Aghanim}, N., {et~al.}
  2014{\natexlab{b}}, \aap, 571, A12

\bibitem[{{Planck Collaboration} {et~al.}(2014{\natexlab{c}}){Planck
  Collaboration}, {Ade}, {Aghanim}, {Alves}, {Arnaud}, {Atrio-Barandela},
  {Aumont}, {Baccigalupi}, {Banday}, {Barreiro}, {Battaner}, {Benabed},
  {Benoit-L{\'e}vy}, {Bernard}, {Bersanelli}, {Bielewicz}, {Bobin}, {Bonaldi},
  {Bond}, {Borrill}, {Bouchet}, {Boulanger}, {Burigana}, {Cardoso}, {Casassus},
  {Catalano}, {Chamballu}, {Chen}, {Chiang}, {Chiang}, {Christensen},
  {Clements}, {Colombi}, {Colombo}, {Couchot}, {Crill}, {Cuttaia}, {Danese},
  {Davies}, {Davis}, {de Bernardis}, {de Rosa}, {de Zotti}, {Delabrouille},
  {D{\'e}sert}, {Dickinson}, {}, {Diego}, {Donzelli}, {Dor{\'e}}, {Dupac},
  {En{\ss}lin}, {Eriksen}, {Finelli}, {Forni}, {Franceschi}, {Galeotta},
  {Ganga}, {G{\'e}nova-Santos}, {Ghosh}, {Giard}, {Gonz{\'a}lez-Nuevo},
  {G{\'o}rski}, {Gregorio}, {Gruppuso}, {Hansen}, {Harrison}, {Helou},
  {Hern{\'a}ndez-Monteagudo}, {Hildebrandt}, {Hivon}, {Hobson}, {Hornstrup},
  {Jaffe}, {Jaffe}, {Jones}, {Keih{\"a}nen}, {Keskitalo}, {Kneissl}, {Knoche},
  {Kunz}, {Kurki-Suonio}, {L{\"a}hteenm{\"a}ki}, {Lamarre}, {Lasenby},
  {Lawrence}, {Leonardi}, {Liguori}, {Lilje}, {Linden-V{\o}rnle},
  {L{\'o}pez-Caniego}, {Mac{\'{\i}}as-P{\'e}rez}, {Maffei}, {Maino},
  {Mandolesi}, {Marshall}, {Martin}, {Mart{\'{\i}}nez-Gonz{\'a}lez}, {Masi},
  {Massardi}, {Matarrese}, {Mazzotta}, {Meinhold}, {Melchiorri}, {Mendes},
  {Mennella}, {Migliaccio}, {Miville-Desch{\^e}nes}, {Moneti}, {Montier},
  {Morgante}, {Mortlock}, {Munshi}, {Naselsky}, {Nati}, {Natoli},
  {N{\o}rgaard-Nielsen}, {Noviello}, {Novikov}, {Novikov}, {Oxborrow},
  {Pagano}, {Pajot}, {Paladini}, {Paoletti}, {Patanchon}, {Pearson}, {Peel},
  {Perdereau}, {Perrotta}, {Piacentini}, {Piat}, {Pierpaoli}, {Pietrobon},
  {Plaszczynski}, {Pointecouteau}, {Polenta}, {Ponthieu}, {Popa}, {Pratt},
  {Prunet}, {Puget}, {Rachen}, {Rebolo}, {Reich}, {Reinecke}, {Remazeilles},
  {Renault}, {Ricciardi}, {Riller}, {Ristorcelli}, {Rocha}, {Rosset},
  {Roudier}, {Rubi{\~n}o-Mart{\'{\i}}n}, {Rusholme}, {Sandri}, {Savini},
  {Scott}, {Spencer}, {Stolyarov}, {Sutton}, {Suur-Uski}, {Sygnet}, {Tauber},
  {Tavagnacco}, {Terenzi}, {Tibbs}, {Toffolatti}, {Tomasi}, {Tristram},
  {Tucci}, {Valenziano}, {Valiviita}, {Van Tent}, {Varis}, {Verstraete},
  {Vielva}, {Villa}, {Wandelt}, {Watson}, {Wilkinson}, {Ysard}, {Yvon},
  {Zacchei}, \& {Zonca}}]{Planck_Int_XV}
---. 2014{\natexlab{c}}, \aap, 565, A103

\bibitem[{{Planck Collaboration} {et~al.}(2015){Planck Collaboration}, {Ade},
  {Alves}, {Aniano}, {Armitage-Caplan}, {Arnaud}, {Atrio-Barandela}, {Aumont},
  {Baccigalupi}, {Banday}, {Barreiro}, {Battaner}, {Benabed},
  {Benoit-L{\'e}vy}, {Bernard}, {Bersanelli}, {Bielewicz}, {Bock}, {Bond},
  {Borrill}, {Bouchet}, {Boulanger}, {Burigana}, {Cardoso}, {Catalano},
  {Chamballu}, {Chiang}, {Colombo}, {Combet}, {Couchot}, {Coulais}, {Crill},
  {Curto}, {Cuttaia}, {Danese}, {Davies}, {Davis}, {de Bernardis}, {de Zotti},
  {Delabrouille}, {D{\'e}sert}, {Dickinson}, {Diego}, {Donzelli}, {Dor{\'e}},
  {Douspis}, {Dunkley}, {Dupac}, {En{\ss}lin}, {Eriksen}, {Falgarone},
  {Finelli}, {Forni}, {Frailis}, {Fraisse}, {Franceschi}, {Galeotta}, {Ganga},
  {Ghosh}, {Giard}, {Gonz{\'a}lez-Nuevo}, {G{\'o}rski}, {Gregorio}, {Gruppuso},
  {Guillet}, {Hansen}, {Harrison}, {Helou}, {Hern{\'a}ndez-Monteagudo},
  {Hildebrandt}, {Hivon}, {Hobson}, {Holmes}, {Hornstrup}, {Jaffe}, {Jaffe},
  {Jones}, {Keih{\"a}nen}, {Keskitalo}, {Kisner}, {Kneissl}, {Knoche}, {Kunz},
  {Kurki-Suonio}, {Lagache}, {Lamarre}, {Lasenby}, {Lawrence}, {Leahy},
  {Leonardi}, {Levrier}, {Liguori}, {Lilje}, {Linden-V{\o}rnle},
  {L{\'o}pez-Caniego}, {Lubin}, {Mac{\'{\i}}as-P{\'e}rez}, {Maffei},
  {Magalh{\~a}es}, {Maino}, {Mandolesi}, {Maris}, {Marshall}, {Martin},
  {Mart{\'{\i}}nez-Gonz{\'a}lez}, {Masi}, {Matarrese}, {Mazzotta},
  {Melchiorri}, {Mendes}, {Mennella}, {Migliaccio}, {Miville-Desch{\^e}nes},
  {Moneti}, {Montier}, {Morgante}, {Mortlock}, {Munshi}, {Murphy}, {Naselsky},
  {Nati}, {Natoli}, {Netterfield}, {Noviello}, {Novikov}, {Novikov},
  {Oppermann}, {Oxborrow}, {Pagano}, {Pajot}, {Paoletti}, {Pasian},
  {Perdereau}, {Perotto}, {Perrotta}, {Piacentini}, {Pietrobon},
  {Plaszczynski}, {Pointecouteau}, {Polenta}, {Popa}, {Pratt}, {Rachen},
  {Reach}, {Reinecke}, {Remazeilles}, {Renault}, {Ricciardi}, {Riller},
  {Ristorcelli}, {Rocha}, {Rosset}, {Roudier}, {Rubi{\~n}o-Mart{\'{\i}}n},
  {Rusholme}, {Salerno}, {Sandri}, {Savini}, {Scott}, {Spencer}, {Stolyarov},
  {Stompor}, {Sudiwala}, {Sutton}, {Suur-Uski}, {Sygnet}, {Tauber}, {Terenzi},
  {Toffolatti}, {Tomasi}, {Tristram}, {Tucci}, {Valenziano}, {Valiviita}, {Van
  Tent}, {Vielva}, {Villa}, {Wandelt}, {Zacchei}, \& {Zonca}}]{Planck_Int_XXII}
{Planck Collaboration}, {Ade}, P.~A.~R., {Alves}, M.~I.~R., {et~al.} 2015,
  \aap, 576, A107

\bibitem[{{Planck Collaboration} {et~al.}(2016{\natexlab{a}}){Planck
  Collaboration}, {Adam}, {Ade}, {Aghanim}, {Alves}, {Arnaud}, {Ashdown},
  {Aumont}, {Baccigalupi}, {Banday}, \& et~al.}]{Planck_2015_X}
{Planck Collaboration}, {Adam}, R., {Ade}, P.~A.~R., {et~al.}
  2016{\natexlab{a}}, \aap, 594, A10

\bibitem[{{Planck Collaboration} {et~al.}(2016{\natexlab{b}}){Planck
  Collaboration}, {Ade}, {Aghanim}, {Alves}, {Arnaud}, {Ashdown}, {Aumont},
  {Baccigalupi}, {Banday}, {Barreiro}, \& et~al.}]{Planck_2015_XXV}
{Planck Collaboration}, {Ade}, P.~A.~R., {Aghanim}, N., {et~al.}
  2016{\natexlab{b}}, \aap, 594, A25

\bibitem[{{Planck Collaboration} {et~al.}(2016{\natexlab{c}}){Planck
  Collaboration}, {Adam}, {Ade}, {Aghanim}, {Arnaud}, {Aumont}, {Baccigalupi},
  {Banday}, {Barreiro}, {Bartlett}, \& et~al.}]{Planck_Int_XXX}
{Planck Collaboration}, {Adam}, R., {Ade}, P.~A.~R., {et~al.}
  2016{\natexlab{c}}, \aap, 586, A133

\bibitem[{{Planck Collaboration} {et~al.}(2017){Planck Collaboration},
  {Aghanim}, {Ashdown}, {Aumont}, {Baccigalupi}, {Ballardini}, {Banday},
  {Barreiro}, {Bartolo}, {Basak}, {Benabed}, {Bernard}, {Bersanelli},
  {Bielewicz}, {Bonaldi}, {Bonavera}, {Bond}, {Borrill}, {Bouchet},
  {Boulanger}, {Bracco}, {Burigana}, {Calabrese}, {Cardoso}, {Chiang},
  {Colombo}, {Combet}, {Comis}, {Crill}, {Curto}, {Cuttaia}, {Davis}, {de
  Bernardis}, {de Rosa}, {de Zotti}, {Delabrouille}, {Delouis}, {Di Valentino},
  {Dickinson}, {Diego}, {Dor{\'e}}, {Douspis}, {Ducout}, {Dupac}, {Dusini},
  {Efstathiou}, {Elsner}, {En{\ss}lin}, {Eriksen}, {Falgarone}, {Fantaye},
  {Finelli}, {Frailis}, {Fraisse}, {Franceschi}, {Frolov}, {Galeotta}, {Galli},
  {Ganga}, {G{\'e}nova-Santos}, {Gerbino}, {Ghosh}, {Giard},
  {Gonz{\'a}lez-Nuevo}, {G{\'o}rski}, {Gregorio}, {Gruppuso}, {Gudmundsson},
  {Hansen}, {Helou}, {Herranz}, {Hivon}, {Huang}, {Jaffe}, {Jones},
  {Keih{\"a}nen}, {Keskitalo}, {Kisner}, {Krachmalnicoff}, {Kunz},
  {Kurki-Suonio}, {Lagache}, {L{\"a}hteenm{\"a}ki}, {Lamarre}, {Lasenby},
  {Lattanzi}, {Lawrence}, {Le Jeune}, {Levrier}, {Liguori}, {Lilje},
  {L{\'o}pez-Caniego}, {Lubin}, {Mac{\'{\i}}as-P{\'e}rez}, {Maggio}, {Maino},
  {Mandolesi}, {Mangilli}, {Maris}, {Martin}, {Mart{\'{\i}}nez-Gonz{\'a}lez},
  {Matarrese}, {Mauri}, {McEwen}, {Melchiorri}, {Mennella}, {Migliaccio},
  {Mitra}, {Miville-Desch{\^e}nes}, {Molinari}, {Moneti}, {Montier},
  {Morgante}, {Moss}, {Naselsky}, {N{\o}rgaard-Nielsen}, {Oxborrow}, {Pagano},
  {Paoletti}, {Partridge}, {Patrizii}, {Perdereau}, {Perotto}, {Pettorino},
  {Piacentini}, {Plaszczynski}, {Polenta}, {Puget}, {Rachen}, {Reinecke},
  {Remazeilles}, {Renzi}, {Rocha}, {Rossetti}, {Roudier},
  {Rubi{\~n}o-Mart{\'{\i}}n}, {Ruiz-Granados}, {Salvati}, {Sandri},
  {Savelainen}, {Scott}, {Sirignano}, {Sirri}, {Stanco}, {Suur-Uski}, {Tauber},
  {Tenti}, {Toffolatti}, {Tomasi}, {Tristram}, {Trombetti}, {Valiviita},
  {Vansyngel}, {Van Tent}, {Vielva}, {Wandelt}, {Wehus}, {Zacchei}, \&
  {Zonca}}]{Planck_Int_L}
{Planck Collaboration}, {Aghanim}, N., {Ashdown}, M., {et~al.} 2017, \aap, 599,
  A51

\bibitem[{{Poh} \& {Dodelson}(2017)}]{Poh+Dodelson_2017}
{Poh}, J., \& {Dodelson}, S. 2017, \prd, 95, 103511

\bibitem[{{Remazeilles} {et~al.}(2016){Remazeilles}, {Dickinson}, {Eriksen}, \&
  {Wehus}}]{Remazeilles+etal_2016}
{Remazeilles}, M., {Dickinson}, C., {Eriksen}, H.~K.~K., \& {Wehus}, I.~K.
  2016, \mnras, 458, 2032

\bibitem[{{Remazeilles} {et~al.}(2017){Remazeilles}, {Banday}, {Baccigalupi},
  {Basak}, {Bonaldi}, {De Zotti}, {Delabrouille}, {Dickinson}, {Eriksen},
  {Errard}, {Fernandez-Cobos}, {Fuskeland}, {Herv{\'{\i}}as-Caimapo},
  {L{\'o}pez-Caniego}, {Martinez-Gonz{\'a}lez}, {Roman}, {Vielva}, {Wehus},
  {Achucarro}, {Ade}, {Allison}, {Ashdown}, {Ballardini}, {Banerji}, {Bartolo},
  {Bartlett}, {Baumann}, {Bersanelli}, {Bonato}, {Borrill}, {Bouchet},
  {Boulanger}, {Brinckmann}, {Bucher}, {Burigana}, {Buzzelli}, {Cai}, {Calvo},
  {Carvalho}, {Castellano}, {Challinor}, {Chluba}, {Clesse}, {Colantoni},
  {Coppolecchia}, {Crook}, {D'Alessandro}, {de Bernardis}, {de Gasperis},
  {Diego}, {Di Valentino}, {Feeney}, {Ferraro}, {Finelli}, {Forastieri},
  {Galli}, {Genova-Santos}, {Gerbino}, {Gonz{\'a}lez-Nuevo}, {Grandis},
  {Greenslade}, {Hagstotz}, {Hanany}, {Handley}, {Hernandez-Monteagudo},
  {Hills}, {Hivon}, {Kiiveri}, {Kisner}, {Kitching}, {Kunz}, {Kurki-Suonio},
  {Lamagna}, {Lasenby}, {Lattanzi}, {Lesgourgues}, {Lewis}, {Liguori},
  {Lindholm}, {Luzzi}, {Maffei}, {Martins}, {Masi}, {McCarthy}, {Melin},
  {Melchiorri}, {Molinari}, {Monfardini}, {Natoli}, {Negrello}, {Notari},
  {Paiella}, {Paoletti}, {Patanchon}, {Piat}, {Pisano}, {Polastri}, {Polenta},
  {Pollo}, {Poulin}, {Quartin}, {Rubino-Martin}, {Salvati}, {Tartari},
  {Tomasi}, {Tramonte}, {Trappe}, {Trombetti}, {Tucker}, {Valiviita}, {Van de
  Weijgaert}, {van Tent}, {Vennin}, {Vittorio}, {Young}, {Zannoni}, \& {for the
  CORE collaboration}}]{Remazeilles+etal_2017}
{Remazeilles}, M., {Banday}, A.~J., {Baccigalupi}, C., {et~al.} 2017, ArXiv
  e-prints, arXiv:1704.04501

\bibitem[{{Siebenmorgen} {et~al.}(2014){Siebenmorgen}, {Voshchinnikov}, \&
  {Bagnulo}}]{Siebenmorgen+Voshchinnikov+Bagnulo_2014}
{Siebenmorgen}, R., {Voshchinnikov}, N.~V., \& {Bagnulo}, S. 2014, \aap, 561,
  A82

\bibitem[{{Smith} {et~al.}(2012){Smith}, {Hanson}, {LoVerde}, {Hirata}, \&
  {Zahn}}]{2012JCAP...06..014S}
{Smith}, K.~M., {Hanson}, D., {LoVerde}, M., {Hirata}, C.~M., \& {Zahn}, O.
  2012, \jcap, 6, 014

\bibitem[{{Stompor} {et~al.}(2016){Stompor}, {Errard}, \&
  {Poletti}}]{Stompor+Errard+Poletti_2016}
{Stompor}, R., {Errard}, J., \& {Poletti}, D. 2016, \prd, 94, 083526

\bibitem[{{Tassis} \& {Pavlidou}(2015)}]{Tassis+Pavlidou_2015}
{Tassis}, K., \& {Pavlidou}, V. 2015, \mnras, 451, L90

\bibitem[{{Thorne} {et~al.}(2017){Thorne}, {Dunkley}, {Alonso}, \&
  {N{\ae}ss}}]{Thorne+etal_2017}
{Thorne}, B., {Dunkley}, J., {Alonso}, D., \& {N{\ae}ss}, S. 2017, \mnras, 469,
  2821

\bibitem[{{Westphal} {et~al.}(2014){Westphal}, {Stroud}, {Bechtel}, {Brenker},
  {Butterworth}, {Flynn}, {Frank}, {Gainsforth}, {Hillier}, {Postberg},
  {Simionovici}, {Sterken}, {Nittler}, {Allen}, {Anderson}, {Ansari}, {Bajt},
  {Bastien}, {Bassim}, {Bridges}, {Brownlee}, {Burchell}, {Burghammer},
  {Changela}, {Cloetens}, {Davis}, {Doll}, {Floss}, {Gr{\"u}n}, {Heck},
  {Hoppe}, {Hudson}, {Huth}, {Kearsley}, {King}, {Lai}, {Leitner}, {Lemelle},
  {Leonard}, {Leroux}, {Lettieri}, {Marchant}, {Ogliore}, {Ong}, {Price},
  {Sandford}, {Tresseras}, {Schmitz}, {Schoonjans}, {Schreiber}, {Silversmit},
  {Sol{\'e}}, {Srama}, {Stadermann}, {Stephan}, {Stodolna}, {Sutton},
  {Trieloff}, {Tsou}, {Tyliszczak}, {Vekemans}, {Vincze}, {Von Korff},
  {Wordsworth}, {Zevin}, {Zolensky}, \& {aff14}}]{Westphal+etal_2014}
{Westphal}, A.~J., {Stroud}, R.~M., {Bechtel}, H.~A., {et~al.} 2014, Science,
  345, 786

\bibitem[{{Zheng} {et~al.}(2017){Zheng}, {Tegmark}, {Dillon}, {Kim}, {Liu},
  {Neben}, {Jonas}, {Reich}, \& {Reich}}]{Zheng+etal_2017}
{Zheng}, H., {Tegmark}, M., {Dillon}, J.~S., {et~al.} 2017, \mnras, 464, 3486

\end{thebibliography}

\end{document}